\documentclass[greek,bibyear]{aa}
\usepackage{times}
\usepackage{amsmath,mathrsfs,amssymb,amsbsy}
\usepackage{graphicx}
\usepackage{subfigure}
\usepackage{paralist}
\usepackage{color}
\usepackage{natbib}
\usepackage[breaklinks=true]{hyperref}
\graphicspath{{plots/}}

\newcommand{\kms}{\,km\,s$^{-1}$} 

\usepackage{gensymb} 
\usepackage{soul} 

\newcommand{\teff}{T$_{\rm eff}$}

\newcommand{\meta}{\hbox{[M/H]}}

\newcommand{\feh}{{\rm {[Fe/H]}}}

\def\afe{\rm [\alpha/Fe]}

\def\kms{\,{\rm km~s^{-1}}}
\def\kpc{\,{\rm kpc}}
\def\mas{\,{\rm mas}}

\def\dex{\,{\rm dex}}

\def\vphi{v_\phi}

\def\vlsr{V_{\rm LSR}}
\def\mgfe{\rm {[Mg/Fe]}}

\def\ltsima{$\; \buildrel < \over \sim \;$}
\def\simlt{\lower.5ex\hbox{\ltsima}}
\def\gtsima{$\; \buildrel > \over \sim \;$}
\def\simgt{\lower.5ex\hbox{\gtsima}}

\def\ltsima{$\; \buildrel < \over \sim \;$}
\def\simlt{\lower.5ex\hbox{\ltsima}}
\def\gtsima{$\; \buildrel > \over \sim \;$}
\def\simgt{\lower.5ex\hbox{\gtsima}}

\begin{document}

\title{The chemodynamics of prograde and retrograde Milky Way stars}
\author{
        Georges~Kordopatis (\selectlanguage{greek}{Gi'wrgos Kordop'aths})\selectlanguage{english},\inst{\ref{oca}}
        Alejandra~Recio-Blanco,\inst{\ref{oca}}
        Mathias~Schultheis,\inst{\ref{oca}}
        Vanessa Hill\inst{\ref{oca}}
        }
\institute{
        Universit\'e C\^ote d'Azur, Observatoire de la C\^ote d'Azur, CNRS, Laboratoire Lagrange, Nice, France \label{oca}
}
        
\abstract
{The accretion history of the Milky Way is still unknown, despite the recent discovery of stellar systems  that stand out in terms of their energy-angular momentum space, such as Gaia-Enceladus-Sausage. In particular, it is still unclear how these groups are linked and to what extent they are well-mixed.}
{We investigate  the similarities and differences in the properties between the prograde and retrograde (counter-rotating) stars and set those results in context by using the properties of Gaia-Enceladus-Sausage, Thamnos/Sequoia, and other suggested accreted populations.}
{We used the stellar metallicities of the major large spectroscopic surveys (APOGEE, Gaia-ESO, GALAH, LAMOST, RAVE, SEGUE) in combination with astrometric and photometric data from Gaia's second data-release. We investigated the presence of radial and vertical metallicity gradients as well as the possible correlations between the azimuthal velocity, $\vphi,$ and metallicity, $\meta,$ as qualitative indicators of the presence of mixed populations.  }
{  We find that a handful of super metal-rich stars exist on retrograde orbits at various distances from the Galactic center and the Galactic plane. We also find that the counter-rotating stars appear to be a well-mixed population, exhibiting radial and vertical metallicity gradients on the order of $\sim -0.04\dex\kpc^{-1}$ and $-0.06\dex\kpc^{-1}$, respectively, with little (if any) variation when different regions of the Galaxy are probed. The prograde stars show a $\vphi-\meta$ relation that flattens -- and, perhaps, even reverses as a function of distance from the plane. Retrograde samples selected to roughly probe Thamnos and Gaia-Enceladus-Sausage  appear to be different populations yet they also appear to be quite linked, as they follow the same trend in terms of the eccentricity versus metallicity space. }
{}

\titlerunning{Prograde and retrograde galactic stars}
\authorrunning{G.~Kordopatis et al.}

\maketitle

\section{Introduction}

The role that accretion events have played in the evolution of the Milky Way and the quest to find possible remnants that are at the origin of the old disc  have been central topics in Galactic archaeology for over than five decades \citep[e.g.][]{Eggen62, Searle78, Gilmore83, Chiba00, Gilmore02,Wyse06, Kordopatis11b}. 
The advent of the Tycho-Gaia Astrometric Solution (TGAS)  catalogue of the European space mission, Gaia \citep{Gaia, GaiaDR1}, and, in particular, the second Gaia data release \citep[GDR2,][]{GaiaDR2}, have enabled us to measure with much greater accuracy the positions and the 3D velocities of millions of stars in a volume  several kiloparsecs wide. Such studies have shed an unprecedented light on the questions cited above. 

On the one hand, the discovery of ripples in the Galactic disc has shown that its morphology and characteristics, at all radii, continue to be impacted by external factors \citep{GaiaKatz18, Antoja18, Laporte18, Laporte19}. On the other hand, the discovery of kinematic groups outside the disc, such as the Gaia-Sausage \citep{Belokurov18, Myeong18b}, Gaia-Enceladus \citep{Helmi18}, Sequoia \citep{Myeong19}, and Thamnos \citep{Koppelman19} are all believed to be remnants of past accretions; however, it is still unclear whether or not they are distinct features and to what extent they might have contributed at the formation of the thick disc or the inner halo \citep[e.g.][]{Haywood18, Fernandez-Alvar19b, Belokurov20}.

In that respect, retrograde stars in the Milky Way hold a key place in our understanding of the assembly history of our Galaxy because there is no clear mechanism that could form them exclusively in situ. Counter-rotating stars in the halo have been discussed since \citet{Majewski92,Carney96, Carollo07, Nissen10, Majewski12}. 
In particular, \citet{Majewski92} identified, via an investigation of proper motions and a multi-colour analysis of a sample of few hundred stars towards the north Galactic pole, a retrograde rotation among stars reaching 5\kpc~from the plane. \citet{Majewski92} reports that they exhibit no radial metallicity gradient, while also stating that this population may be younger, on average, than the dynamically hot metal-poor stars that are closer to the plane.  This result was later confirmed by \citet{Carney96} using a kinematically biased sample of 1500 stars and by \citet{Carollo07} using calibration data from 20\,000 stars from the Sloan Extension for Galactic Understanding and Exploration \citep[SEGUE,][]{Yanny09}. 

\citet{Nissen10} carried out a spectroscopic observation  at high resolution of a sample of 94 kinematically selected dwarf stars  and found that the low-$\afe$ sequence stars identified in the $\afe$ vs $\feh$ plane were mostly counter-rotating targets. According to the authors, some of the low-$\afe$ sequence stars may have originated from  $\omega$\,Centauri's globular cluster  progenitor \citep[the latter also being on a retrograde orbit; see e.g. ][]{Dinescu02}, 
a hypothesis that has also been suggested by  \citet{Majewski12}, using low-resolution ($R\sim2600$) spectra of $\sim3\,000$ stars, and by \citet{Myeong18c}, using a catalogue of $\sim 62\,000$ halo stars from a crossmatch of Gaia DR1, the SDSS DR 9 \citep{Ahn12}, and LAMOST DR 3 \citep{Luo15} catalogues. 

Using the exquisite GDR2 data, \citet{Gaia_Helmi18} found that a non-negligible amount of the halo stars in the extended Solar neighbourhood   that are somewhat less bounded than the Sun are on retrograde orbits. Compared with cosmological simulations, they conclude that this rate of counter-rotating stars, despite being rare, is not unexpected.  
\citet{Helmi18} suggested that they originate from the merger with Gaia-Enceladus, the latter in a slightly counter-rotating orbit with a mass ratio of 4:1 \citep[see also][]{Gallart19, Feuillet20}.

While  the metallicity extent, chemical structure as well as the number of subpopulations that constitute these retrograde stars have already been investigated in previous papers in action-angular momentum space \citep[e.g.][]{Helmi18, Myeong18b, Myeong19, Koppelman19, Feuillet20, Naidu20}, the 
investigation of the correlation between the stellar azimuthal velocity, $\vphi,$  and the stellar metallicity, $\meta$, or iron abundance, $\feh$, can provide an additional insight in this endeavour \citep[e.g.][for the identification of thick disc stars and characterisation of its properties]{Kordopatis13b}. The level of correlation between two or more parameters can also highlight the mechanisms that have formed and shaped a stellar population that is being considered. 

While this was already visible in previous datasets \citep[see e.g.  ][]{Carney96}, \citet{Spagna10, Kordopatis11b, Lee11}  isolated the  correlation between kinematics and metallicity  for thick disc  stars and measured it to be on the order of $\partial \vphi / \partial \meta\approx +50\kms\dex^{-1}$. A mild negative correlation has also been measured for the chemically identified thin disc stars in \citet{Recio-Blanco14, Kordopatis17, Allende-Prieto16}. Whereas for the thin disc, this correlation is well-understood as the effect of blurring and churning in the disc \citep[e.g.][]{Sellwood02, Schonrich09a, Minchev10}, multiple explanations can be found in the literature concerning the origin of the positive correlation for the thick disc stars. Indeed, it has been suggested that it could either be the signature of the collapse of a primitive gas cloud \citep[e.g.][]{Kordopatis17}, a signature of inside-out formation and gas re-distribution in the primitive disc \citep{Schonrich17}, or, as recently suggested  by \citet{Minchev19}, a correlation resulting from the superposition of mono-$\afe$ sub-populations with negative slopes (as in the thin disc). In the latter case, the measured positive correlation is due to the combination of several populations of different ages with different relative weights and proportions (the so-called Yule-Simpson paradox). 

In this paper, we aim to study the chemokinematics of the stars in the Gaia-sphere, with a particular focus on the retrograde targets, in order to identify trends that could shed some light on the formation origins of the thick disc and the origin of those retrograde stars. Section\,\ref{sec:dataset} describes the dataset used in this analysis. Sections\,\ref{sec:gradients} and \ref{sec:vphi_correlations} show the spatial metallicity gradients and the correlations between the azimuthal velocity and metallicity for the prograde and retrograde stars, whereas Sect.\,\ref{sec:Enceladus} discusses selections probing Gaia-Enceladus-Sausage stars and other accreted populations, selected in the phase-space. Finally, Sect.\,\ref{sec:conclusions} presents our conclusions.

\section{Description of the dataset used and the quality cuts applied}
\label{sec:dataset}
We used the distance, velocity, and age catalogue of stars of \citet{Sanders18} to select targets from LAMOST \citep[DR3,][]{Deng12}, APOGEE \citep[DR14,][]{Abolfathi18}, RAVE \citep[DR5,][]{Kunder17}, SEGUE  \citep[DR12,][]{Yanny09}, GALAH \citep[DR2,][]{Buder18}, and Gaia-ESO \citep[DR3,][]{Gilmore12} surveys. 

We applied the quality-flag present in the catalogue to remove all the stars for which the spectroscopic parameters are too far from the isochrones to have reliable distances and ages \citep[i.e. those that do not have the value of \texttt{flag} equal to zero; see][for more details]{Sanders18}. We also removed the RAVE-on \citep{Casey17} entries in order to avoid duplicates with RAVE-DR5. 
As far as the \teff~range is concerned, we only kept the stars between 3500\,K and 6800\,K in order to avoid too cool or too hot stars for which spectra parameterisation is intrinsically difficult to obtain and its results uncertain, thus introducing a potential bias\footnote{A visual inspection of the Kiel diagram confirmed that stars outside this range of \teff~ should indeed be removed due to their abnormal location in this diagram.}. 

In addition to the filters above, further cuts were required in order to make sure our sample contained only single stars. To achieve this, we cross-matched the sample of stars that fulfil the above criteria with the GDR2 archive, based on the GDR2 \texttt{sourceid}. We extracted the renormalised unit weight error (RUWE) of the targets and discarded the stars with a RUWE greater than 1.2 (suggesting that Gaia's astrometric solution has not converged appropriately and that the considered stars are potential binaries)\footnote{See \url{https://gea.esac.esa.int/archive/documentation/GDR2/Gaia_archive/chap_datamodel/sec_dm_main_tables/ssec_dm_ruwe.html}}, an astrometric excess noise greater than 1, and, finally, a parallax relative uncertainty ($\sigma \varpi /\varpi$) greater than 0.1. The latter filter ensures that what dominates the stellar distance estimation is the parallax measurement and not the  prior adopted in \citet{Sanders18}. 
We note that the adopted distances do not take into consideration the zero-point offset reported for GDR2 parallaxes \citep[e.g.][]{Lindegren18, Schonrich19}, as it has a dependence on the position on the sky, the magnitude, and  the colour of the star, which complicates an application  that does not involve the introduction of further biases \citep[e.g.][]{GaiaDR2, Arenou18}. That said, the global effect of this zero point on the velocities and gradients is discussed in the relevant sections below and in greater detail in Appendix~\ref{sec:parallax_offset}.

\begin{figure}
\begin{center}
\includegraphics[width=\linewidth, angle=0]{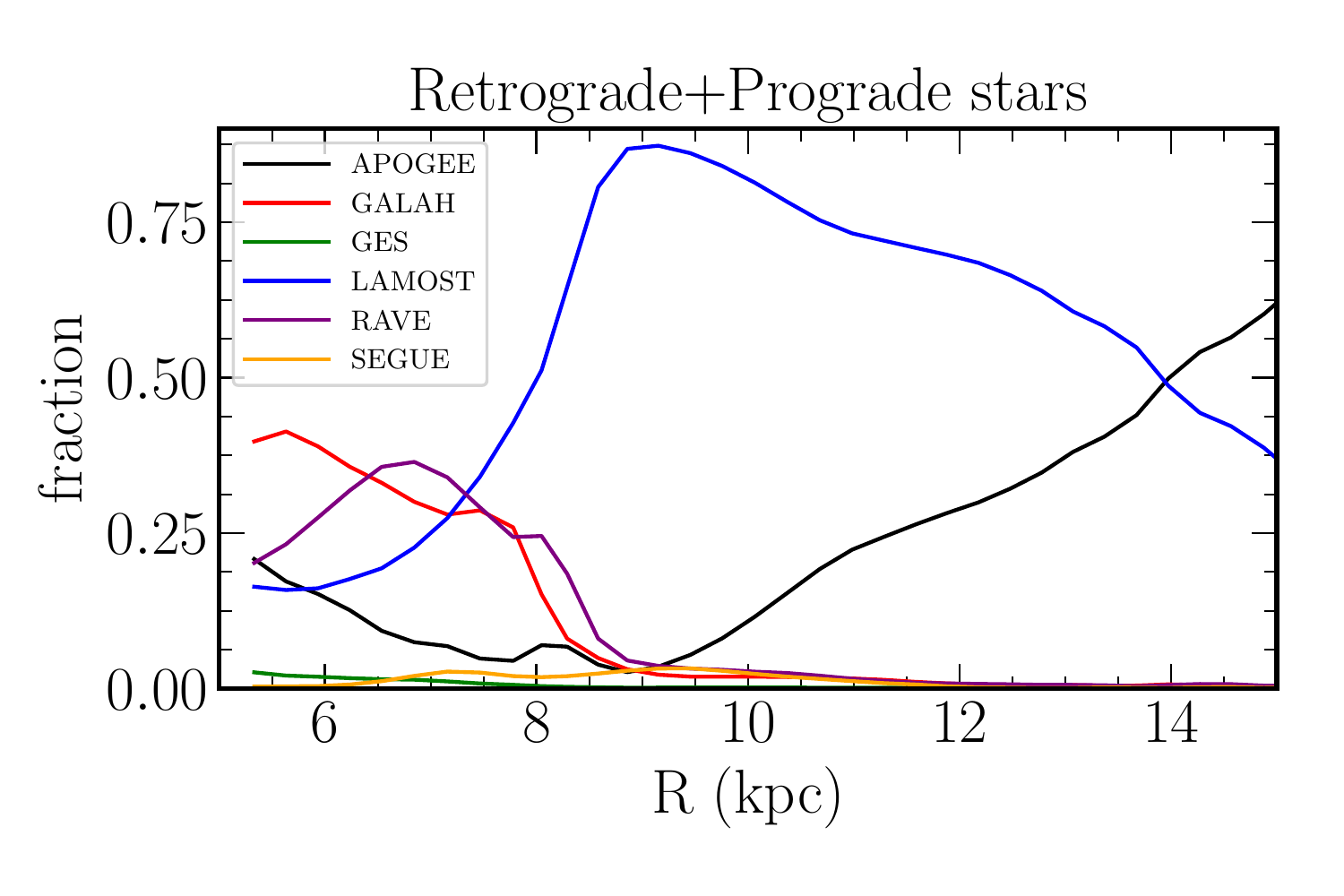}\\
\includegraphics[width=\linewidth, angle=0]{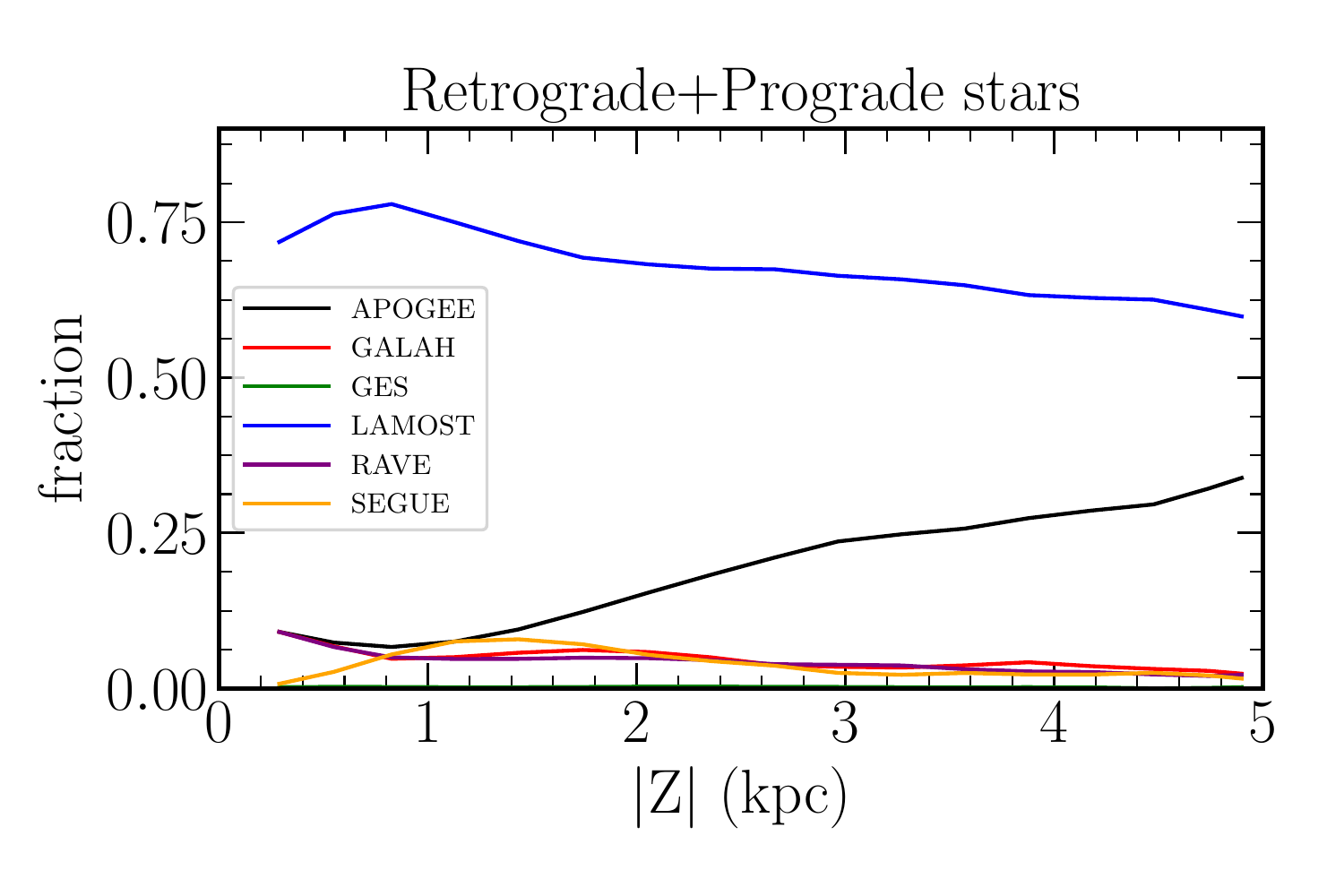}
\caption{Relative fraction of targets within each survey for the stars closer than 5\kpc~from the Galactic plane after application of the quality cuts described in Sect.~\ref{sec:dataset}, as a function of the Galactocentric radius, $R$  (top), and as a function of absolute distance from the Galactic plane, $|Z|$ (bottom). }
\label{fig:R_surveys}
\end{center}
\end{figure}

\begin{figure*}
\begin{center}
\includegraphics[width=\linewidth, angle=0]{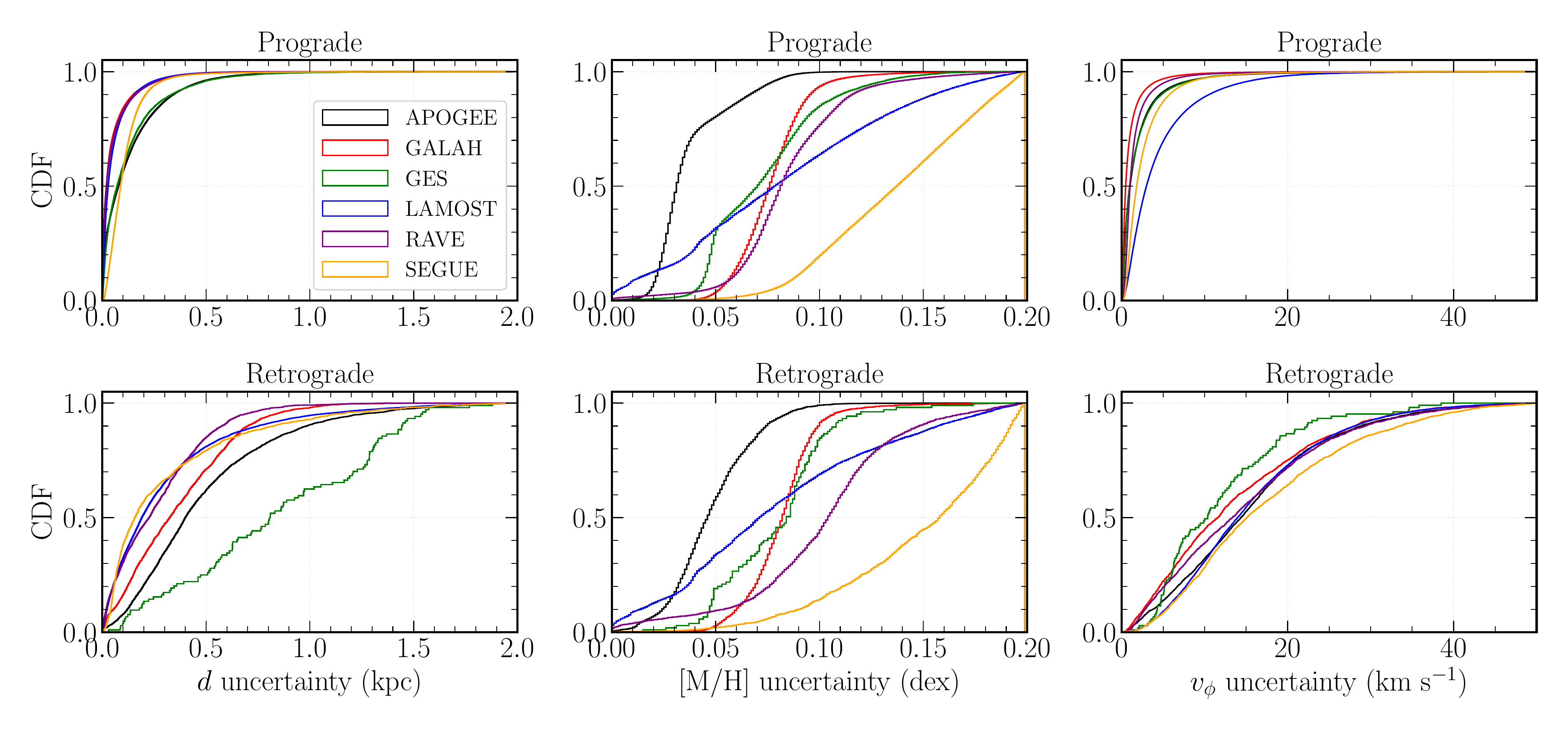}
\caption{Cumulative distribution functions of the uncertainties in line-of-sight distance (left), metallicity (middle), and $\vphi$ (right) split by survey (different colours), and by prograde (top) and retrograde (bottom) populations (as defined in Sect.~\ref{sec:gradients}). }
\label{fig:uncertainties}
\end{center}
\end{figure*}

We also discarded the stars that have an uncertainty in metallicity greater than 0.2\dex, a $\vphi$ uncertainty greater than $50\kms$, and a Galactocentric cartesian $X$ position that suggests they are located past the Galactic center (in order to avoid probing entirely different regions of the Galaxy at a given $R$). 
Our final working sample was obtained by removing the inter-survey repeats, using the \texttt{duplicate} keyword in the  \citet{Sanders18} table, which preferentially selects  the stars  in the order: APOGEE, GALAH, GES, RAVE, LAMOST, and SEGUE (for a comparison between the metallicity, the distance, and the azimuthal velocity for the repeated inter-survey stars, see Appendix~\ref{sec:intersurveyComparison}). Eventually, we ended up with  2\,419\,655 unique stars, out of the 4\,906\,746 entries in the initial catalogue. The relative fractions of each catalogue as a function of the Galactocentric radius, $R$, and absolute distance from the Galactic plane, $|Z|$, are shown in Fig.~\ref{fig:R_surveys}.  The cumulative distribution functions of the uncertainties in distance, metallicity and $\vphi$, split by individual surveys, are shown in Fig.~\ref{fig:uncertainties}. As anticipated, this figure shows that low-resolution surveys tend to have larger uncertainties in metallicity, and, to some extent, $\vphi$, yet the uncertainties in distance are mostly driven by the apparent magnitude of the targets. 

The stellar orbits were computed using the \texttt{galpy} code \citep{Bovy_galpy} with  the \texttt{MWPotential2014} and the action-angle formalism for axisymmetric potentials using \citet{Binney12a}'s Staeckel approximation. 
To be compatible with \citet{Sanders18} velocities and priors, we adopted the Solar peculiar velocity from \citet{Schonrich10}, the velocity of the local standard of rest as $\vlsr=240\kms$, and the Sun's position equal to $(R,Z)_\odot=(8.2,0.015)\kpc$. 

\begin{figure}
\begin{center}
\includegraphics[width=\linewidth, angle=0]{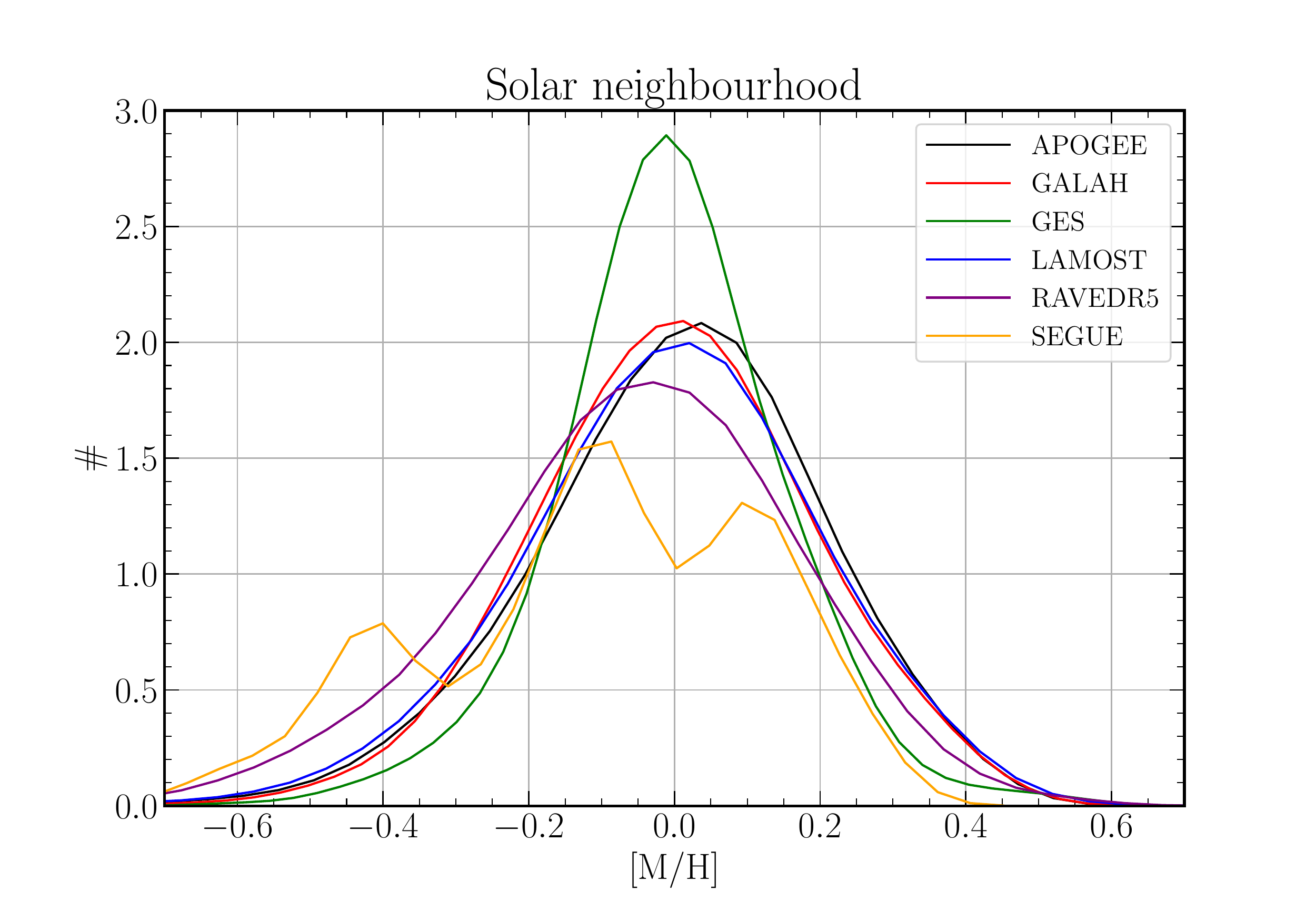}
\caption{ Normalised metallicity distributions (obtained using a kernel density estimation with an Epanechnikov kernel and a smooting parameter of 0.06 dex) for stars close to the Sun ($|R_\odot -R|<0.2\kpc, |Z|<0.2\kpc$), colour-coded by survey.  The plot has been truncated at -0.7, in order to better visualise the peak of the distributions. A good agreement is found between the different surveys in the sense that they are all peaking at the same value ($\sim0$). We note that the SEGUE distribution contains less than 100 stars (most of them K dwarfs), resulting to a noisier shape. }
\label{fig:Solar_neigborhoood_mdf}
\end{center}
\end{figure}

Once we had our final dataset in hand, we verified again that the metallicities of the stars belonging to the different surveys are roughly on the same scale. This was achieved by visually inspecting that the metallicity distributions close to the Sun's position ($|R_\odot -R|<0.2\kpc, |Z|<0.2\kpc$) peaked at the same value. The metallicity distributions, shown in Fig.~\ref{fig:Solar_neigborhoood_mdf}, exhibit a good agreement given the distinct selection functions and the discrepancies already identified in Fig.~\ref{fig:survey_offsets}. 

\section{Radial and vertical metallicity gradients}
\label{sec:gradients}

\begin{figure*}
\begin{center}
\includegraphics[width=\linewidth, angle=0]{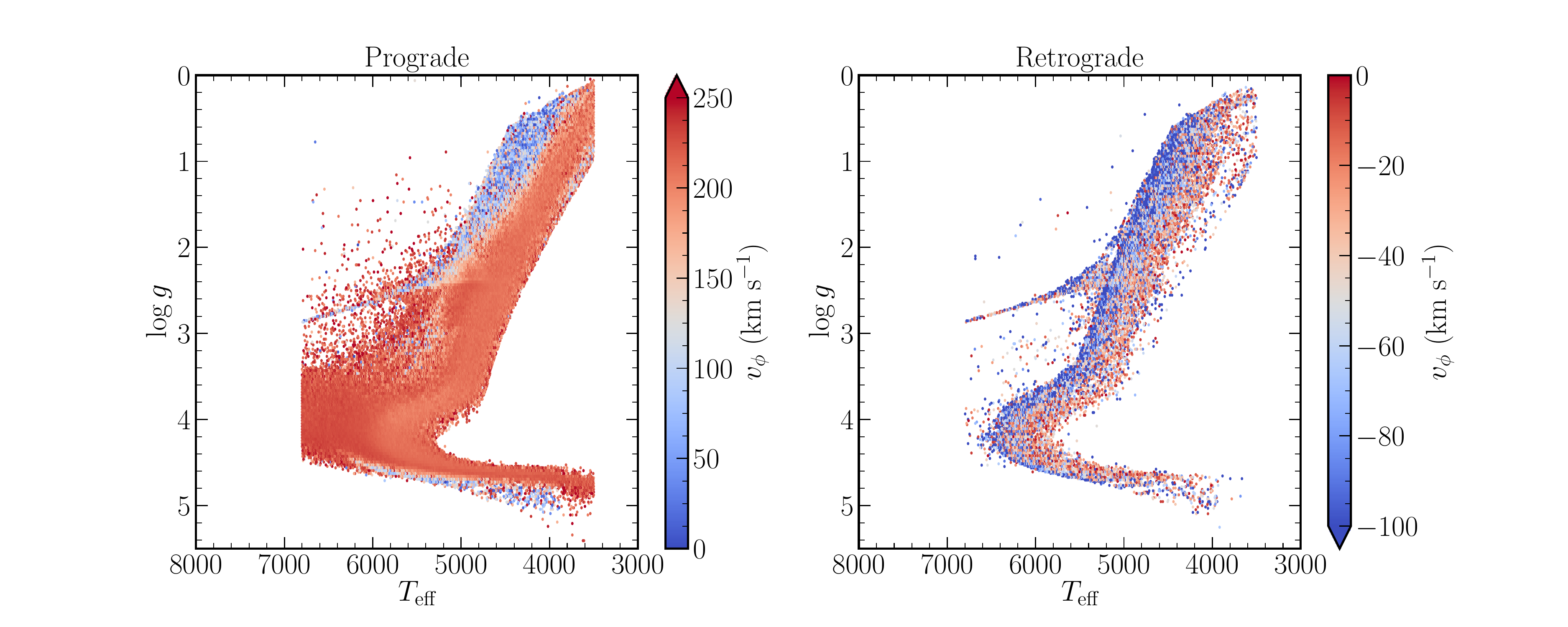}
\caption{Kiel diagrams for the stars that fulfilled our quality criteria for the prograde (left) and retrograde (right) populations, colour-coded by azimuthal velocity $\vphi$. }
\label{fig:Kiel_diagrams}
\end{center}
\end{figure*}

Figure~\ref{fig:Kiel_diagrams} shows the Kiel diagram of the  
prograde 
($\vphi>0\kms$, 2\,397\,183 stars) 
and retrograde 
($\vphi<0\kms$, 22\,472 stars) samples.
Simply by comparing the two diagrams, especially at the turn-off and red giant branch (RGB) regions, we can already  notice that the age range of the retrograde stars is smaller than that of the prograde stars, yet the large width of the turn-off as well as the width of the RGB, suggests that retrograde stars encompass a range of ages.

\begin{figure}
\begin{center}
\includegraphics[width=\linewidth, angle=0]{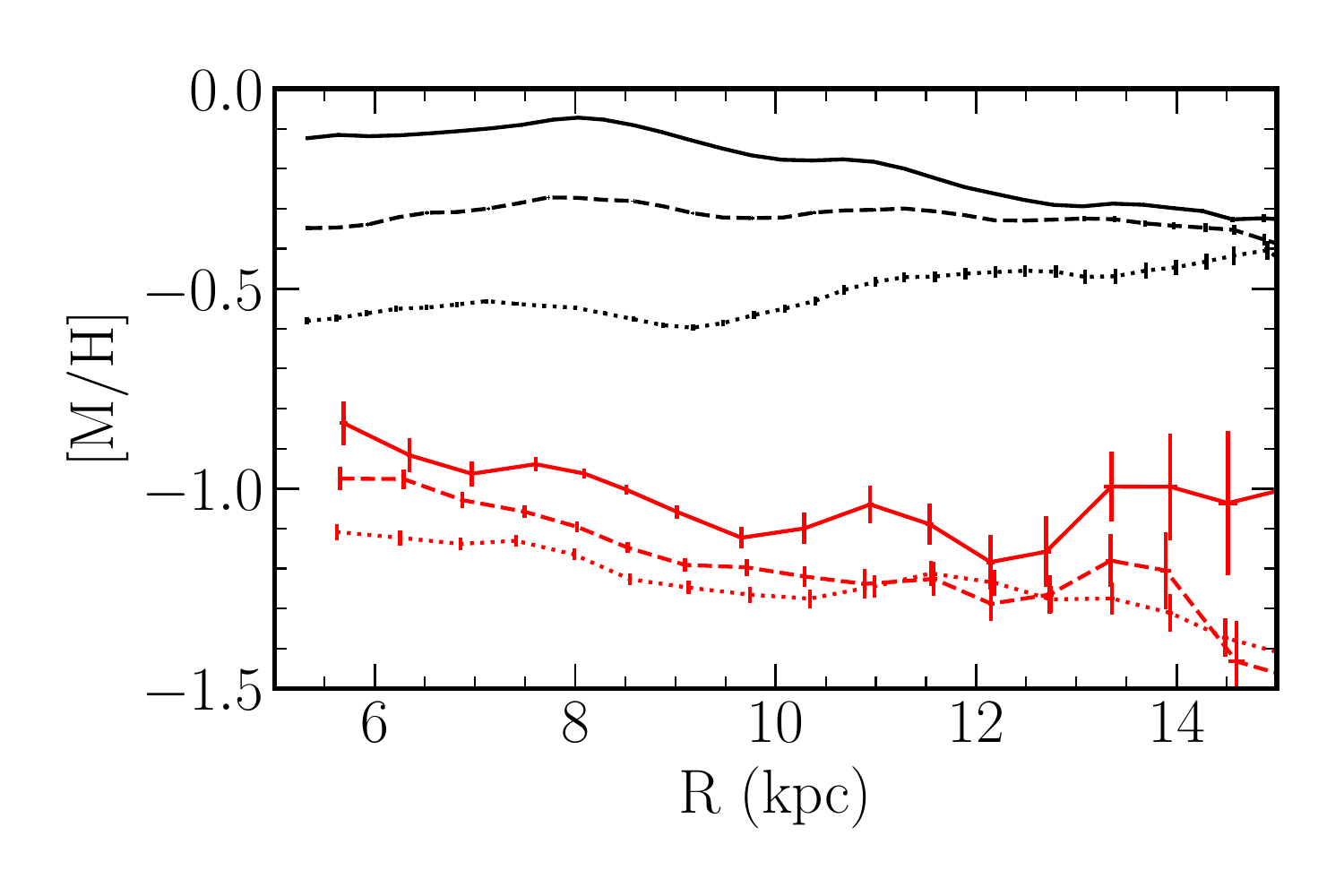}
\caption{Radial gradients for the prograde stars (in black) and the retrograde stars (in red). Solid, dashed, and dotted lines correspond to selections for $0.2 \leq |Z| < 1\kpc$, $1 \leq |Z| < 2\kpc$,  and $2 \leq |Z| < 4.5\kpc$, respectively. Error bars are computed as $\sigma/\sqrt{N}$. }
\label{fig:radial_gradients}
\end{center}
\end{figure}

\begin{figure}
\begin{center}
\includegraphics[width=\linewidth, angle=0]{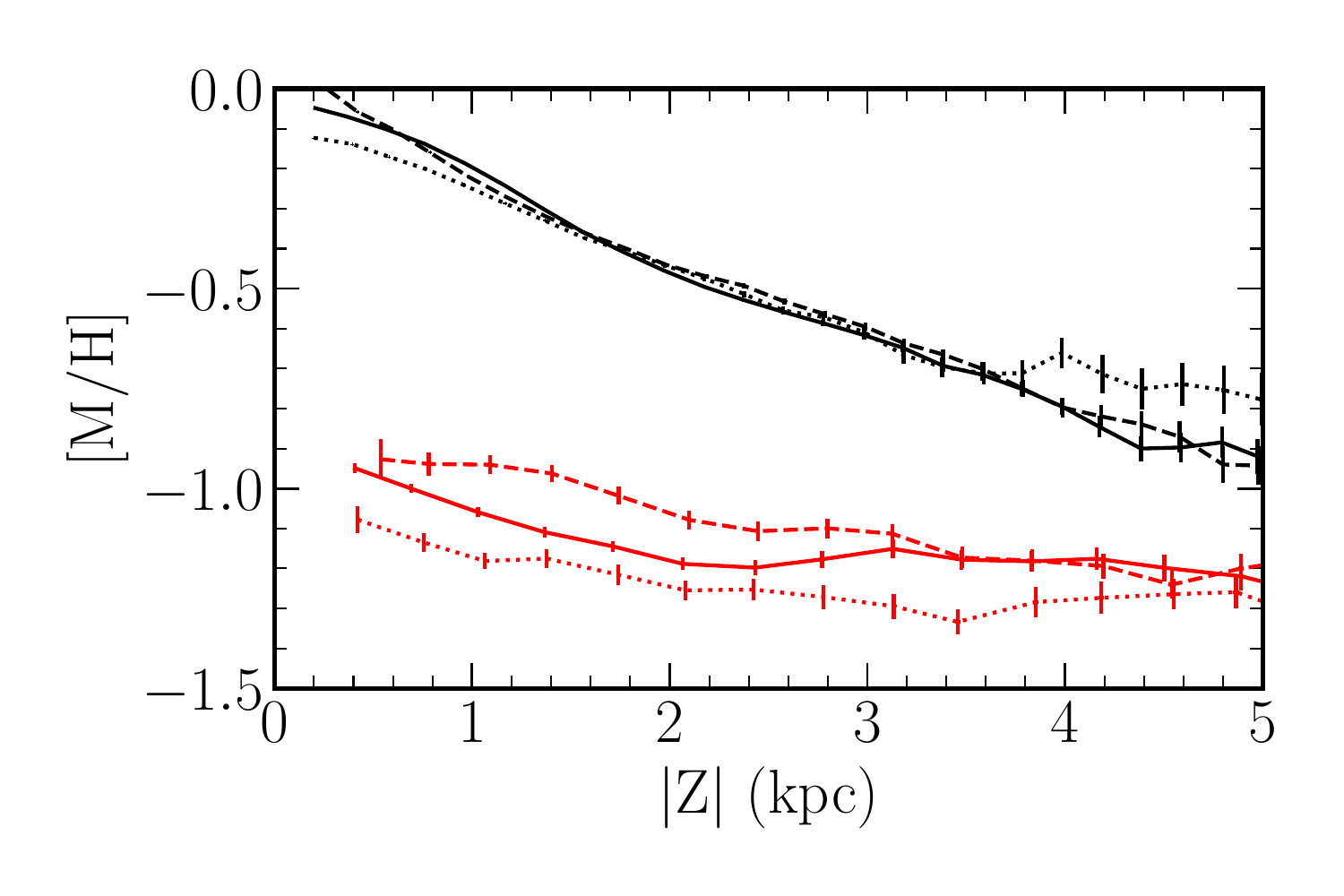}
\caption{Vertical gradients for the prograde stars (in black) and the retrograde stars (in red). Solid, dashed, and dotted lines correspond to selections for $7.2 \leq R < 8.2\kpc$, $5.2 \leq R < 7.2\kpc$,  and $9.2 \leq R < 11.2\kpc$, respectively.  }
\label{fig:vertical_gradients}
\end{center}
\end{figure}

Figures~\ref{fig:radial_gradients} and \ref{fig:vertical_gradients} show the radial and vertical gradients for the two populations for selected ranges of distances from the plane and Galactocentric radii. The associated measured gradients are reported in Tables\,\ref{tab:Radial_Gradients} and \ref{tab:Vertical_Gradients}, respectively, and are discussed in the following subsections.

\begin{figure}
\begin{center}
\includegraphics[width=\linewidth, angle=0]{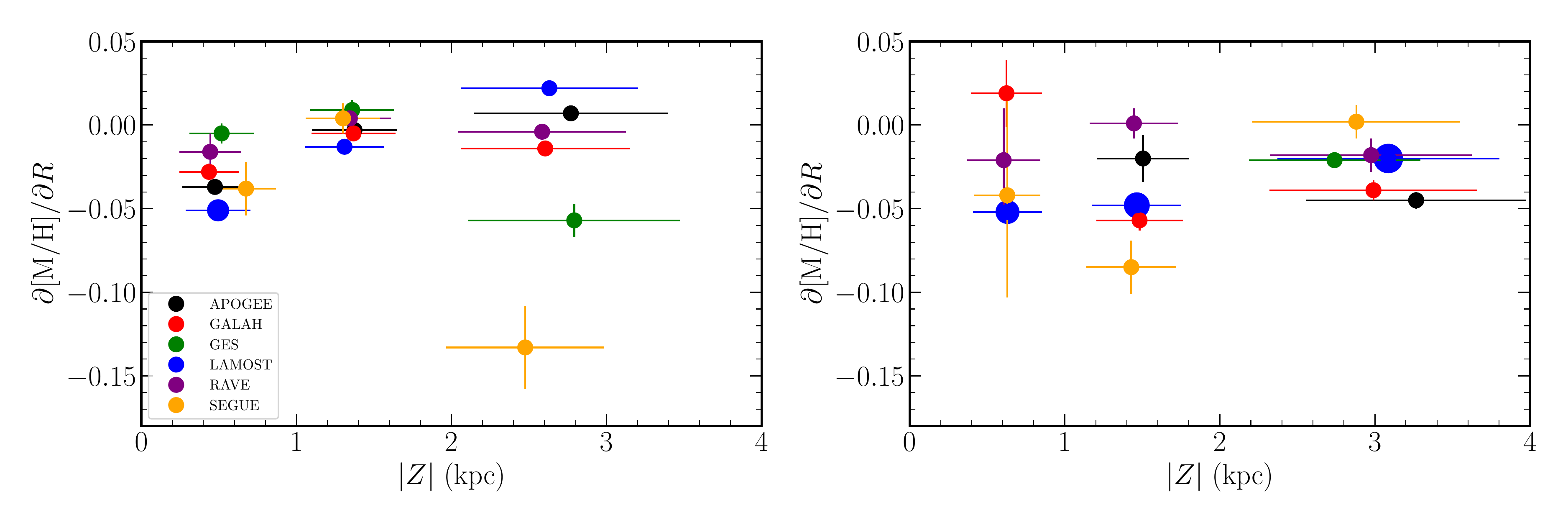}\\
\includegraphics[width=\linewidth, angle=0]{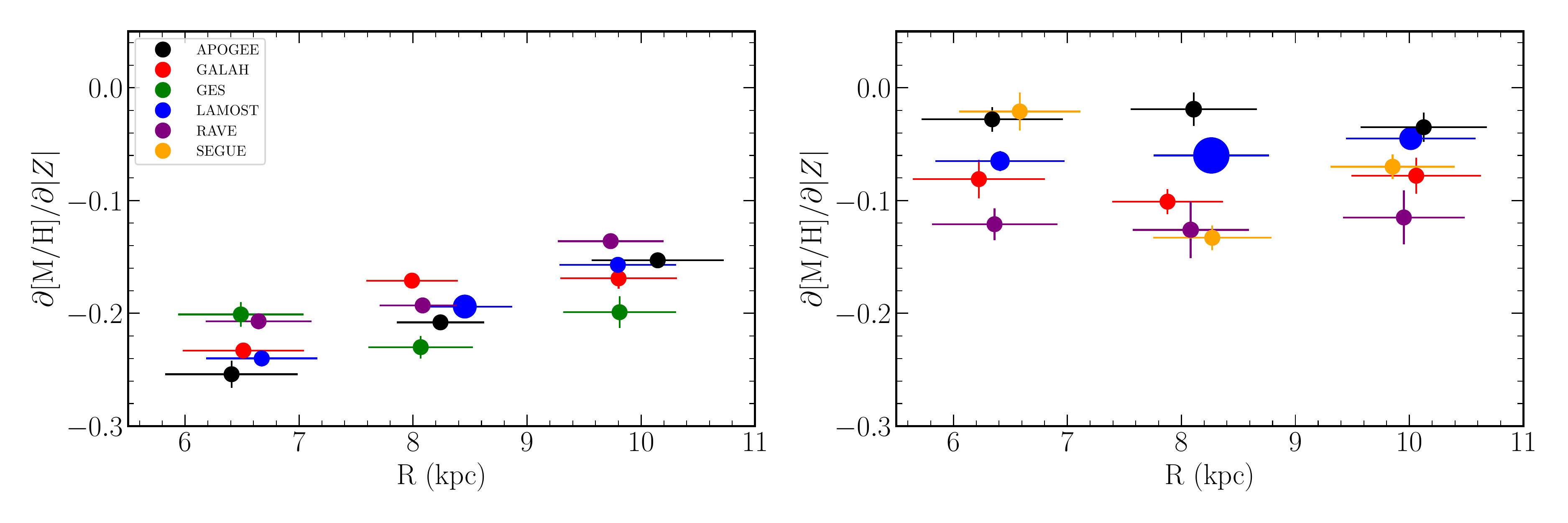}
\caption{Radial (top) and vertical (bottom) metallicity gradients for the prograde (left) and retrograde (right) samples, split by survey. The vertical error bars correspond to the uncertainty of the fit of the gradient, whereas the horizontal error bar represents the dispersion in $|Z|$ or $R$, respectively, where the gradient was measured. The size of the dots is a visual aid proportional (within each panel) to the number of stars each survey contains in the considered distance-bin.}
\label{fig:gradients_survey}
\end{center}
\end{figure}

\subsection{Metallicity gradients for the prograde stars}
We find that the radial metallicity gradient of the prograde stars exhibits two regimes. First, in the inner disc ($5 \leq R\leq 8\kpc$), a globally null gradient is present, at all $|Z|$ ($0.01\pm0.002 \dex\kpc^{-1}$ closest to the plane\footnote{The positive gradient for the stars closest to the plane is due to the fact that closer to Sun, the average distance of the stars from the Galactic plane is smaller (due to the footprint on the sky of the surveys, trying to avoid the Galactic plane).}, $-0.005 \pm 0.005\dex\kpc^{-1}$ in the $Z=[1-2]$\kpc~bin). 

For Galactocentric distances greater than $8\kpc$, the metallicity gradient flattens as one moves further from the plane, regardless of whether  we correct for the zero-point offset on the parallaxes (see also Tables\,\ref{tab:Radial_Gradients_zeropoint} and \ref{tab:Vertical_Gradients_zeropoint}). For the distances derived without the zero-point correction, the metallicity gradient  eventually reverts from a negative one to a slightly positive one at distances where the thick disc dominates (i.e. $|Z|>2\kpc$). 
This result, first reported in \citet{Boeche13} using RAVE-DR4 data \citep{Kordopatis13b},  is not due to small number statistics (as approximately $5 \cdot 10^4$ targets are still available at those distances) nor to the non-homogeneity of the considered samples, as the metallicity gradient can also be measured when using LAMOST, APOGEE, RAVE-DR5, and GALAH separately (see top-left panel of Fig.~\ref{fig:gradients_survey}). 
The inversion  of the gradient (or the flattening) can be interpreted as a thick disc that is more centrally concentrated and more metal-poor than the thin disc combined with a thin disc that exhibits a flare at large radii. This result has also been suggested in studies in which the discs have been defined chemically ($\afe$-high population for the thick disc and $\afe$-low population for the thin disc), such as in  \citet{Bensby11, Hayden15, Kordopatis15b, Minchev17, Anders17}.

\begin{table}
\caption{Measured Galactocentric radial gradients in the $R=[5-15]\kpc$ range.}
\label{tab:Radial_Gradients}
\begin{center}
\begin{tabular}{ccc}
\hline \hline
$|Z|$-range & Prograde & Retrograde \\
(kpc) & (dex/kpc) & (dex/kpc) \\ \hline
$[0.2-1.0]$&  $-0.037\pm 0.003$ & $-0.041\pm 0.008$ \\ 
$[1.0-2.0]$& $-0.004 \pm  0.002$ & $-0.048 \pm 0.005$ \\ 
$[2.0-4.5]$& $0.012 \pm 0.003$ & $-0.028\pm 0.003$\\ \hline
 \end{tabular}
 
 \end{center}
 \end{table}

 \begin{table}
\caption{Measured vertical gradients relative to the Galactic plane in the range $|Z|=[0-5]\kpc$}
\label{tab:Vertical_Gradients}
\begin{center}
\begin{tabular}{ccc}
\hline \hline
$R$-range & Prograde & Retrograde \\
(kpc) & (dex/kpc) & (dex/kpc) \\ \hline
$[5.2-7.2]$ & $-0.236  \pm 0.007$ & $-0.071 \pm  0.007$ \\
$[7.2-9.2]$& $-0.202 \pm  0.006$ & $-0.062 \pm 0.009$ \\
$[9.2-11.2]$ & $-0.171  \pm 0.005$ & $-0.043  \pm 0.007$ \\ \hline

 \end{tabular}
 \end{center}
 \end{table}
 
 Similar to the radial metallicity gradients that exhibit a vertical dependency, we find that the vertical metallicity gradients of the prograde stars also present a radial dependence, even when individually considering  the stars belonging to each survey (see bottom left panel of Fig.~\ref{fig:gradients_survey}). The gradients we measure for all of the selected stars range from $-0.24\dex\kpc^{-1}$ at the inner disc to $-0.17\dex\kpc^{-1}$ at the outer disc.   The vertical gradients that are derived for the range $5.3\leq R \leq 9.2\kpc$ are similar to the gradients previously found in the literature, on the order of $-0.2\dex\kpc^{-1}$ to $-0.27\dex\kpc^{-1}$  \citep[see][for a comparison between the different surveys]{Nandakumar17}, compatible with a mixture of two populations of different scale-heights, namely, $h_{\rm z, thin}\sim300$\,pc and $h_{\rm z, thick}\sim1000$\,pc \citep[e.g.][]{Gilmore83}.

\subsection{Metallicity gradients for the retrograde stars}
\label{subsec:retrograde_gradients}

As far as the trends of the retrograde stars are concerned, they are strikingly different than the ones found for the prograde stars. 
The radial gradients are always significantly negative, on the order of $\sim-0.04\dex\kpc^{-1}$, with no clear indication of a dependency with $|Z|$, at least up to $|Z|\sim2\kpc$, (even when the surveys are considered individually).

As far as the vertical gradients are concerned, these are much shallower than the ones derived for the prograde stars, yet they are still significant, on the order of $\sim-0.05\dex\kpc^{-1}$  (see also bottom-right panel of Fig.~\ref{fig:gradients_survey} for values derived from each survey separately).  
A mild dependence on the radius might exist, when considering all of the stars simultaneously, in the sense that the outer radial  bin seems to have a slightly flatter vertical gradient than the one measured at the inner Galaxy, regardless of the zero-point parallax correction. 
 
 The absence, or small dependence, of the vertical metallicity gradient variations on the distance from the Galactic center and of the radial metallicity gradient on the distance from the Galactic plane is compatible with either a well-mixed population or a single population (coming from e.g. a single accretion) that would have a pre-existent metallicity gradient \citep[e.g.][]{Abadi03}. 
Further investigation is therefore needed in order to better understand this retrograde population. In the next section, we scrutinise the correlations between $\vphi$ and the metallicity.

\section{Correlations between rotational velocity and metallicity}
\label{sec:vphi_correlations}
\begin{figure*}
\begin{center}
\includegraphics[width=0.9\linewidth, angle=0]{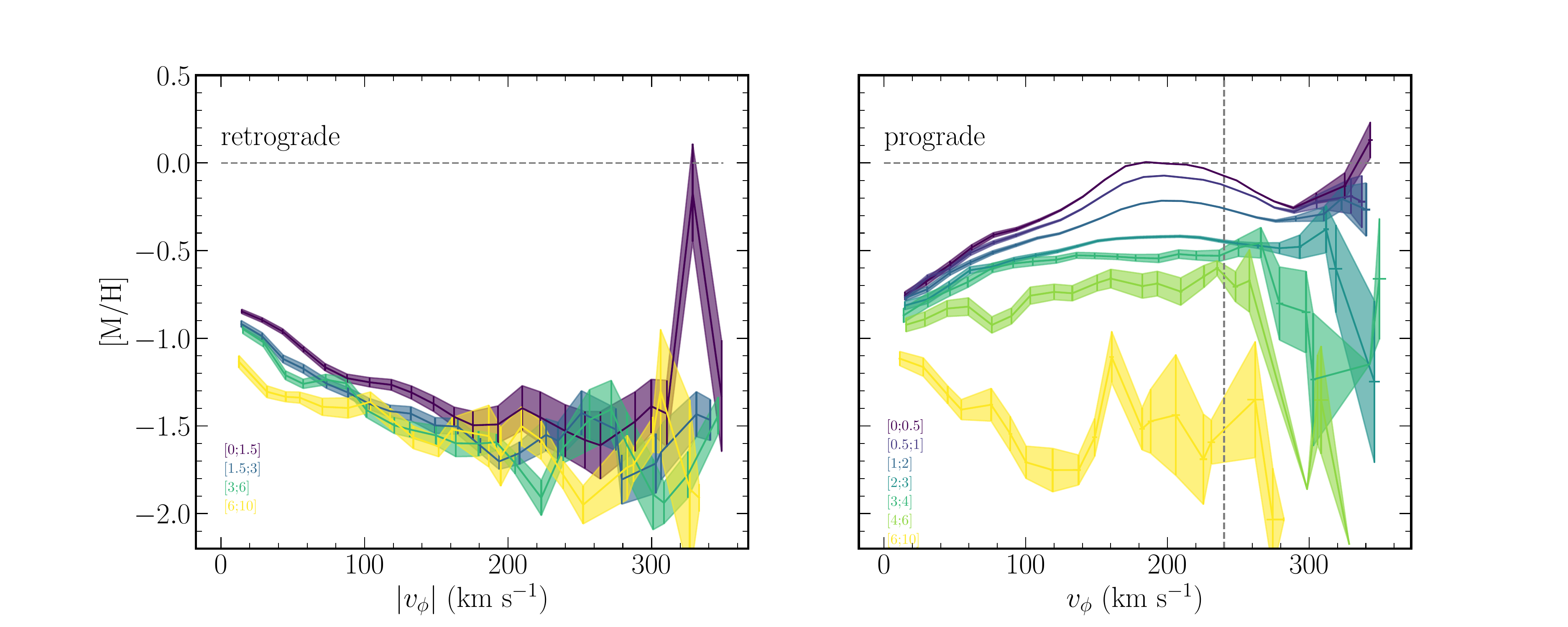}
\caption{$\vphi$ vs $\meta$ for the stars in the Solar cylinder ($R=[7.2-9.2]\kpc$). Different colours correspond to different distances from the Galactic plane (darker colours correspond to the closest to the plane, yellow colours to the farthest distances), the range in $\kpc$ being reported at the lower-left corner of each plot. The vertical dashed line is at $\vlsr$ and horizontal dashed line is at solar metallicity. Vertical error bars correspond to $\sigma_{\meta} / \sqrt{N}$, where $N$ is the number of stars in a given bin and $\sigma_{\meta}$ is the standard deviation of the metallicity inside that bin. }
\label{fig:vphi_metal_sun}
\end{center}
\end{figure*}

\begin{figure*}
\begin{center}
\includegraphics[width=0.9\linewidth, angle=0]{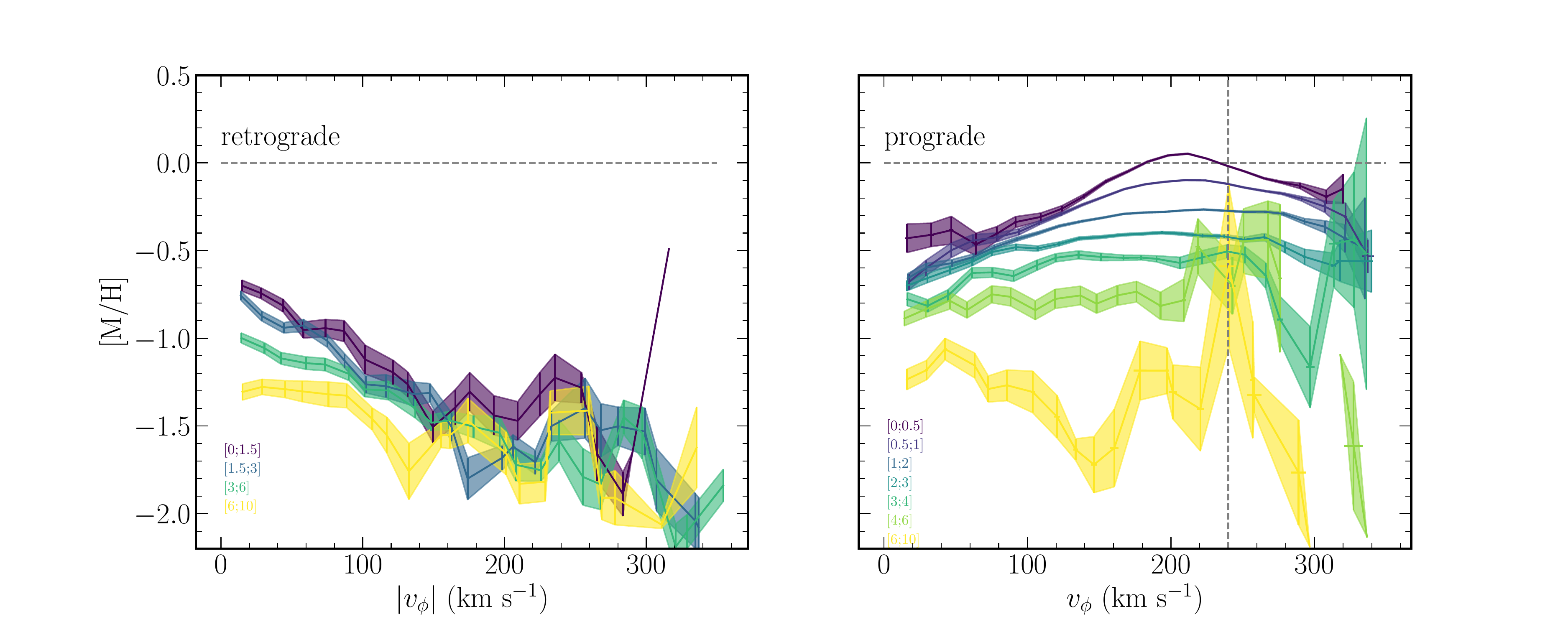}
\caption{Same as Fig.~\ref{fig:vphi_metal_sun}, but for the inner Galaxy ($R=[5.2-7.2]\kpc$).}
\label{fig:vphi_metal_inner}
\end{center}
\end{figure*}

\begin{figure*}
\begin{center}
\includegraphics[width=0.9\linewidth, angle=0]{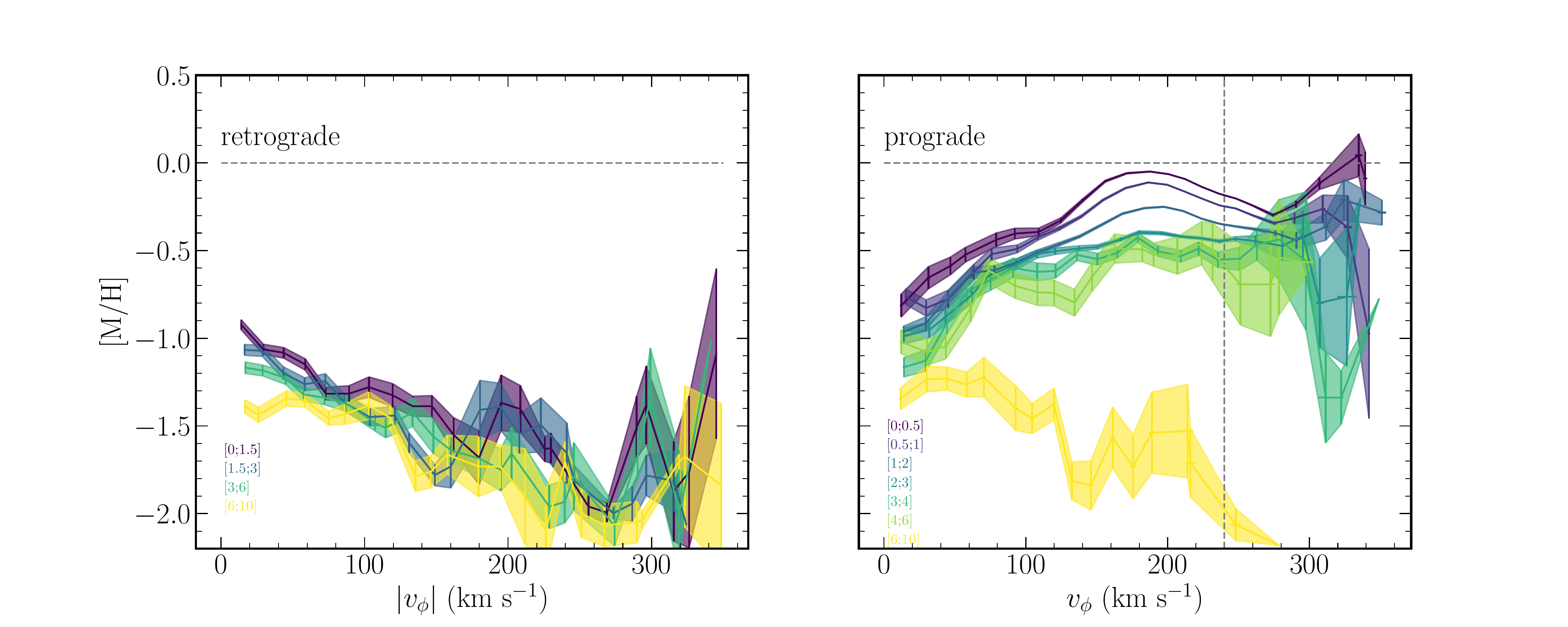}
\caption{Same as Fig.~\ref{fig:vphi_metal_sun}, but for the outer Galaxy ($R=[9.2-11.2]\kpc$).}
\label{fig:vphi_metal_outer}
\end{center}
\end{figure*}

Figures\,\ref{fig:vphi_metal_sun}, \ref{fig:vphi_metal_inner}, and \ref{fig:vphi_metal_outer} show, for the Solar neighbourhood ($R=[7.2-9.2]\kpc$), the inner Galaxy ($R=[5.2-7.2]\kpc$), and the outer Galaxy ($R=[9.2-11.2]\kpc$), respectively,  the median $\meta$ for $30\kms$-wide bins in $\vphi$, where each $\vphi$-bin overlaps by half a step with the previous one.  The prograde stars are shown on the right-hand side  panels and the retrograde stars on the left-hand side panels.  We note that these trends do not change (although they become more noisy due to fewer stars) when only one survey is taken into account at once \citep[which is in agreement with][who state that the selection functions  of the considered surveys do not  significantly alter the measured Galactic gradients]{Nandakumar17}, nor when the zero-point offset in parallax is corrected for.

We also note that the usual way  of plotting the chemokinematic correlations that can be found in the literature is  the opposite than the one presented in these figures: commonly, it is $\vphi$ that is marginalised over metallicity-bins. However, by doing so, the tails of the velocity distribution are systematically missed or ignored, which is not what we are looking to do in this study. For this reason, the plots that follow are not obtained in the ``usual" way, that is, using $\vphi=f(\meta)$.

\subsection{Trends for the prograde stars}
Regarding the prograde stars, we can find the expected and known trends in the disc: 
close to the plane (purple colours on the right panels of Figs.~\ref{fig:vphi_metal_sun} to \ref{fig:vphi_metal_outer}), stars show a negative correlation down to $\vphi\sim180\kms$ (the typical velocity-lag of the thick disc). Then they show a positive correlation for lower angular momenta. 

The negative correlation, typically found for the chemically defined thin disc stars \citep[e.g.][]{Lee11,Recio-Blanco14, Allende-Prieto16, Kordopatis17} is due to the blurring of the stellar orbits: older thin disc stars have a higher eccentricity than young stars via more Lindblad resonances with the spiral arms; this eventually allows them to visit radii that are far from their birth radius \citep{Lynden-Bell72}. As a consequence, old outer thin disc stars can reach the solar neighbourhood at their pericentre, hence with velocities lower than the local standard of rest (LSR), and the inner thin disc stars can reach the solar neighbourhood at their apocentre, hence with velocities higher than the LSR. Because of the radial metallicity gradient in the thin disc, stars that are more metal-rich than the locally born stars tend to move faster than the LSR, and the stars that are more metal-poor than the locally born stars will tend to move slower than the LSR. Comparing the metallicities at which the inflexion happens in the Figs.~\ref{fig:vphi_metal_sun}, \ref{fig:vphi_metal_inner}, and \ref{fig:vphi_metal_outer}, we can notice that it shifts from super-solar metallicities at the inner disc to sub-solar metallicities at our outer disc sample, which is as expected for a disc with a negative radial metallicity gradient.

The positive correlation that we find for stars with $\vphi\lesssim 160-180 \kms$ \citep[typically found for the thick disc stars, see][]{Spagna10, Kordopatis11b, Lee11, Recio-Blanco14, Kordopatis17, ReFiorentin19} is often used as an argument against radial migration in the thick disc. We note, however, that \citet{Minchev19} suggested that this correlation may be due to the superposition of stars of different ages, each mono-age population itself having  a negative correlation (the so-called Yule-Simpson effect). The precision of the ages available for our sample does not allow us to either support or reject this statement. 

That said, we find that the trends become flatter as we move farther from the plane. Eventually, for $6<|Z|<10\kpc$ (or even $|Z|>4\kpc,$ when the zero-point offset in parallax is taken into account), that is, where the canonical thick disc still represents a significant fraction of the stars (assuming a scale-height of $1\kpc$ and a local normalisation of $\sim 0.1$), the trends seem to become negative for all of the stars over all the $\vphi$-range (i.e. even at large velocities where the ratio canonical thick disc or canonical halo is high) and at all Galactocentric radii. This result is compatible with \citet{Minchev19} and could potentially suggest that the thick disc stars far from the plane and the inner halo might be one single mono-age population. To our knowledge, this flattening trend has not been identified in any previous study.

\subsection{Retrograde stars}
Unlike the prograde stars, all of the retrograde stars seem to exhibit similar behaviour at any distance from the plane and any distance from the Galactic centre. The correlation is always positive, on the order of $\sim25-30\kms\dex^{-1}$, again suggesting that the probed population  is well-mixed at all $R$ (one would otherwise expect different trends). 
However, the retrograde sample contains, rather unexpectedly, super-solar metallicity (SMR) stars (see for instance  Figs.~\ref{fig:metal_vphi_all_R} and \ref{fig:metal_vphi_all_Z} as well as Fig.~\ref{fig:vphi_metal_sun}, left panel, around $\vphi\sim-325\kms$). Although only $\sim600$ of them ($\sim 300$ when the zero-point offset in parallax is taken into account, observed mostly by the APOGEE, LAMOST, and RAVE surveys), they show an inverse (i.e. negative) correlation with metallicity, similar to the thin disc prograde stars. These  SMR stars are located at all Galactic sky coordinates $(\ell,b)$,  with 50 per cent of them being closer than $1\kpc$ from the Galactic plane, and reaching distances up to  $|Z|\sim6-7\kpc$.
They are also seen in both the inner and the outer Galaxy (up  to $R\sim 15\kpc$). Interestingly, when  the zero-point offset in parallaxes is not taken into account, 20\% of the retrograde stars are located in the bulge region, that is, between $1<R<2.5\kpc$,  but the latter sample becomes prograde  when correcting for the zero point (also see the plots in Figs.~\ref{fig:metal_vphi_all_R} and \ref{fig:metal_vphi_all_Z}). 

\begin{figure}
\begin{center}
\includegraphics[width=1.15\linewidth, angle=0]{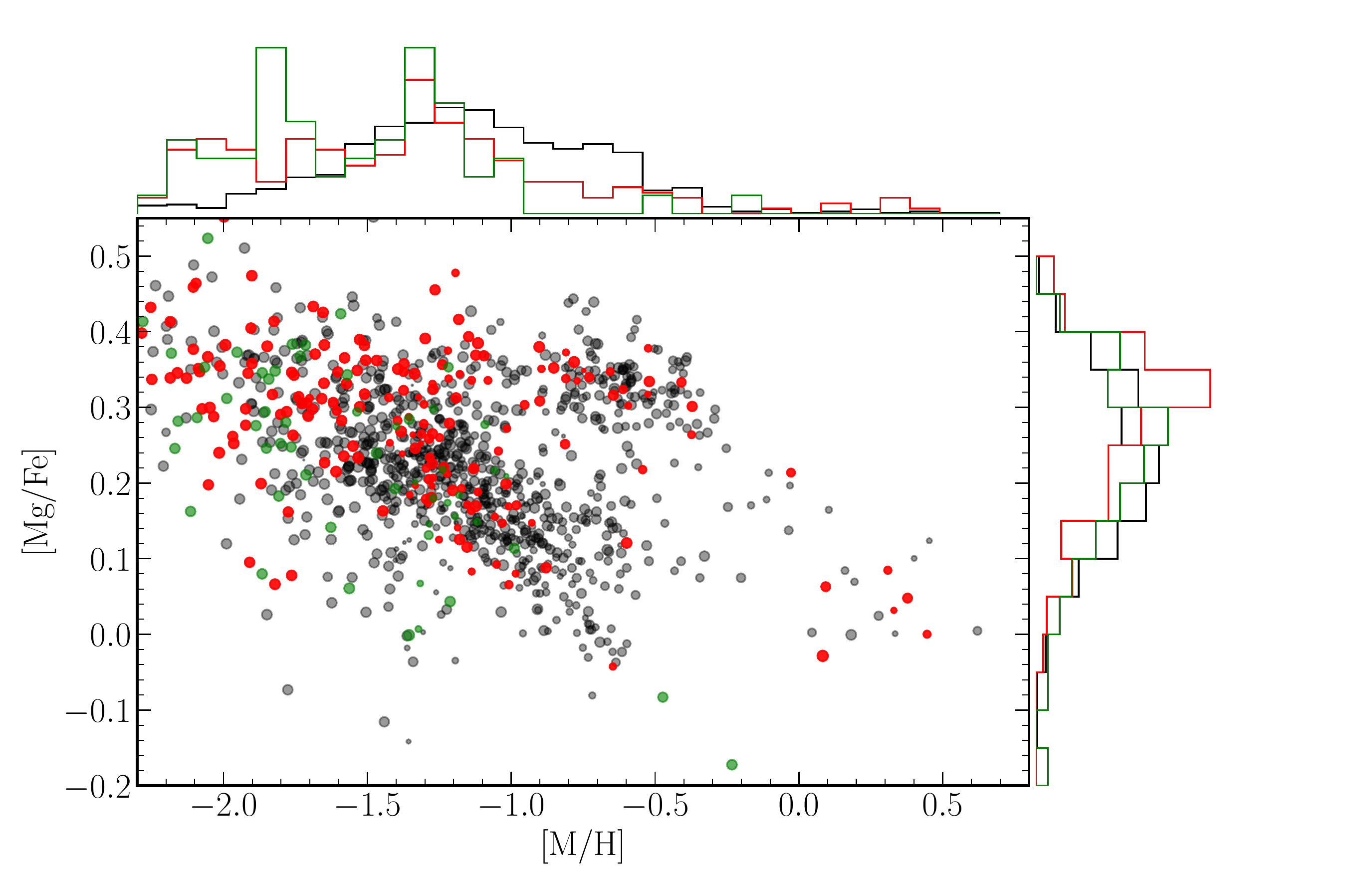}
\caption{Scatter plot of magnesium abundance as a function of metallicity for the retrograde stars in the APOGEE-DR16 catalogue and marginalised normalised histograms. Black, red, and green points and histograms correspond to  counter-rotating stars that we labelled `Gaia-Enceladus-Sausage', `Thamnos', and `other', respectively  (the criteria for selecting those stars are described in Sect.~\ref{sec:Enceladus}). The sizes of the points are indicative and correspond to the estimated ages from \citet[][]{Sanders18};  we note, however, that the majority of these stars are giants and, hence, their ages are unreliable.  }
\label{fig:CR_alphas}
\end{center}
\end{figure}

\begin{figure*}
\begin{center}
\includegraphics[width=\linewidth, angle=0]{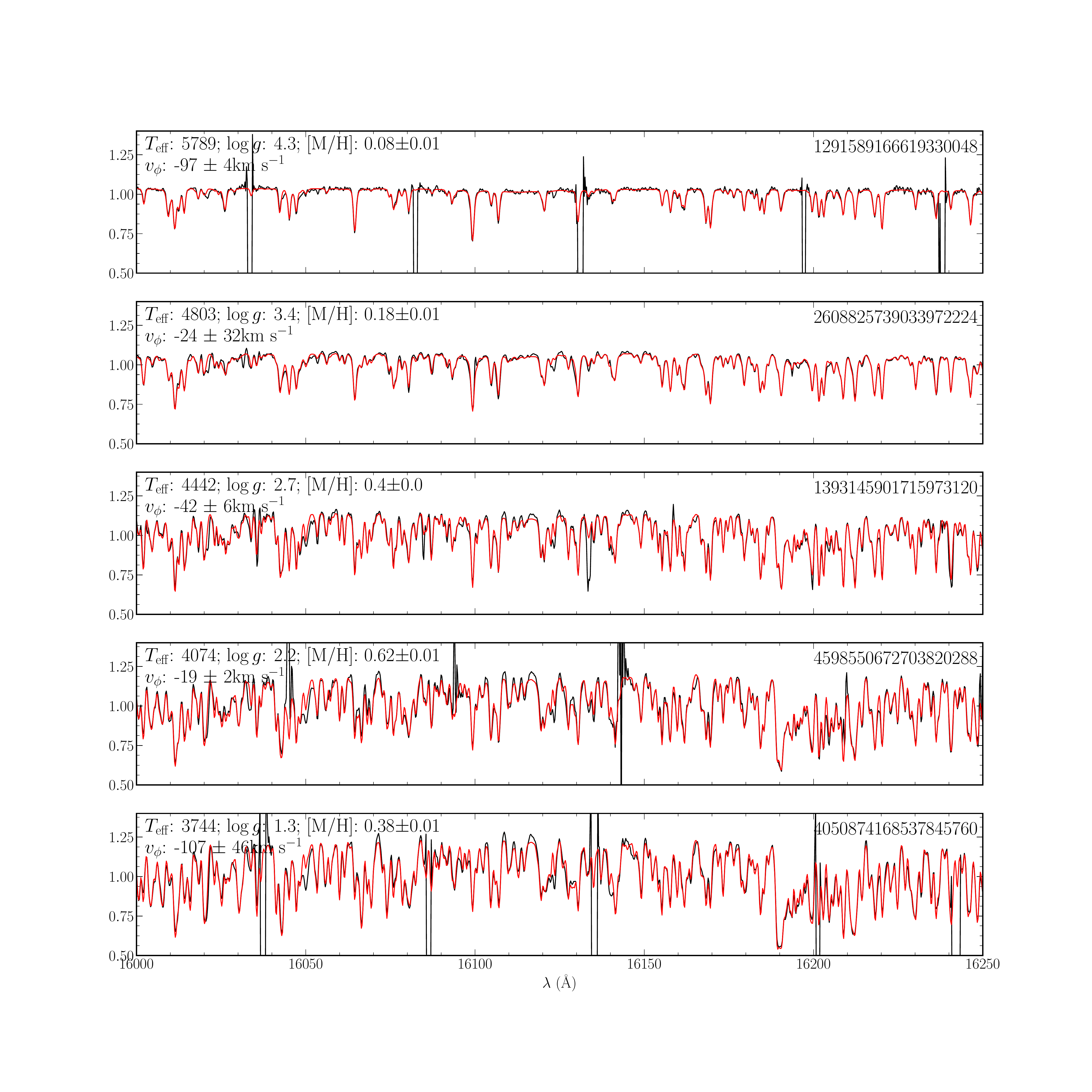}
\caption{APOGEE DR16 spectra (in a selected arbitrary wavelength range, in black) and best-fit templates (in red) for five of the super-solar metallicity retrograde stars in our sample. The Gaia \texttt{sourceid} is reported at the top-right corner and the atmospheric parameters of the stars, as well as their azimuthal velocity, $\vphi,$ at the top-left corner.  }
\label{fig:SMR_spectra}
\end{center}
\end{figure*}

We cross-matched our retrograde sample with the APOGEE-DR16 catalogue \citep[][]{Ahumada20} and out of the 1197 stars which  have a reliable abundance determination\footnote{Throughout the paper, we remove from the APOGEE DR16 catalogue those stars with a signal-to-noise ratio (S/N) lower than 60, as well as those flagged with the following keywords: \texttt{BAD\_PIXELS, BAD\_RV\_COMBINATION, LOW\_SNR, PERSIST\_HIGH, SUSPECT\_BROAD\_LINES, VERY\_BRIGHT\_NEIGHBOR, NO\_ASPCAP\_RESULT, STAR\_BAD, SN\_BAD, STAR\_BAD, CHI2\_BAD, COLORTE\_BAD}.}, we find that 18 have $\meta>0$ (see Fig.~\ref{fig:CR_alphas})\footnote{When taking into account the  zero point of the parallax, we end up with 475 stars, amongst which five are SMR.}.  A specific investigation of the stellar spectra belonging to the APOGEE, LAMOST and RAVE surveys will be performed in future papers with the aim of confirming whether those SMR retrograde stars are indeed as metal-rich as suggested by the the \citet{Sanders18} catalogue.  We show, nevertheless, in Fig.~\ref{fig:SMR_spectra},  the APOGEE DR16 spectra and their best-fit templates for five of those SMR retrograde targets as a preview of the study that will follow, also indicating that these stars have a good parameterisation.

Recently, \citet{Fragkoudi20}, \citet{Belokurov20}  and \citet{Grand20} revived the idea that the thick disc may have formed  after a violent gas-rich merger \citep[e.g.][]{Brook04}. In this context, because of the compressible nature of the gas, some of the newborn stars could have been formed on counter-rotating or radial orbits. The time and mass of this gas-rich merger, according to \citet{Grand20}, could be then inferred from the properties and the fraction of the locally born retrograde stars. 
Figure~\ref{fig:CR_alphas} shows the magnesium abundance as a function of metallicity for all of the retrograde stars in the APOGEE DR16 sample. 
We can see from this figure two chemical sequences starting at $\meta\gtrsim-1.5$: a high-$\alpha$, typically associated with stars born in situ, and a low-$\alpha$, associated with accreted populations \citep[e.g.][]{Nissen10}. Whereas the majority of the counter-rotating stars are located in the low$-\alpha$ sequence, a non-negligible fraction of them, extending up to super-solar metallicities, are on the high-$\alpha$ sequence, indicating that they are possibly born locally. As we do not know the selection function of our sample, we cannot draw conclusions on their relative proportions. However, the fact that they are slightly $\alpha$-enhanced, even for $\meta>0$, corroborates to the fact that  they were likely born from in situ material.

According to \citet{Helmi20, Naidu20}, the  largest fraction of  retrograde low-$\alpha$ stars  belong to Gaia-Enceladus-Sausage, especially at large distances from the plane. In the scenario where Gaia-Enceladus-Sausage was a massive merger (with a mass-ratio of 4:1), this could imply that such a massive galaxy would have an internal metallicity gradient and puffed-up the Galactic disc that was present at the time of the merger. The low-metallicity accreted stars from such a merger would therefore be deposed first, on high eccentricity,  and the metal-rich stars (formed close to the center of the accreted galaxy) would be deposited on more circular orbits due to dynamical friction \citep[e.g.][]{Koppelman20}.  In the next section, we  investigate more closely the case of Gaia-Enceladus-Sausage and try to identify differences with the other retrograde stars and groups of accreted stars, as identified recently in the literature.

\section{Gaia-Enceladus-Sausage, Thamnos/Sequoia, $\omega$Cen, and the other counter-rotating stars}
\label{sec:Enceladus}
In what follows, we define three sub-samples of counter-rotating stars. 

First, we labelled, as `Gaia-Enceladus-Sausage' the targets fulfilling the quality criteria presented in Sect.~\ref{sec:dataset} and having $L_Z<0\kpc\kms$ and $\vphi>-200\kms$ as well as $e>0.65$ 
(16\,283 out of the 22\,472 counter-rotating stars, LAMOST and APOGEE targets encompassing 56.4 and 16.1 per cent of the targets, respectively). This selection is somewhat compatible with remnants of a counter-rotating merger of initial inclination of approximately $30\deg$ \citep[see ][Sect.\,4.2.1]{Helmi20} 
but we stress that it ignores the prograde counter-part of the merger remnants and is likely contaminated by other accreted populations \citep[see for example][]{Feuillet20}. Therefore, our sample should not be directly compared with other Gaia-Enceladus or Gaia-Sausage samples in the literature.

We also selected stars probing the chemically and dynamically peculiar population identified by \citet[][]{Koppelman19} and dubbed \emph{`Thamnos'},  the targets  fulfilling the previous quality criteria and having $L_Z<0\kpc\kms$, $e<0.65,$ as well as $\vphi>-200\kms$ 
(4\,007 stars, LAMOST and APOGEE  encompassing 52.1 and 19.2 per cent of the targets, respectively), 
that is, the stars having similar angular momentum and $\vphi$ as our `Gaia-Enceladus-Sausage' sample, but having low eccentricities. 
 Despite  being still unclear if `Thamnos' is  part of the low-energy tail of Sequoia \citep[see][]{Myeong19, Naidu20}, in what follows we nevertheless labelled this sample  `Thamnos'  as the majority of Sequoia stars are on higher energy orbits.



Finally, all the other retrograde stars  that are not `Gaia-Enceladus-Sausage' nor `Thamnos', with $L_z<0\kpc\kms$, are labelled in what follows as "other" (1\,461 stars, LAMOST, RAVE and APOGEE encompassing 58.1, 13.6, and  13.1 per cent of the targets, respectively). It is noteworthy that this category comprises,  `Sequoia' stars as well as other sub-populations \citep[see e.g.][]{Naidu20}.


\begin{figure}
\begin{center}
\includegraphics[width=\linewidth, angle=0]{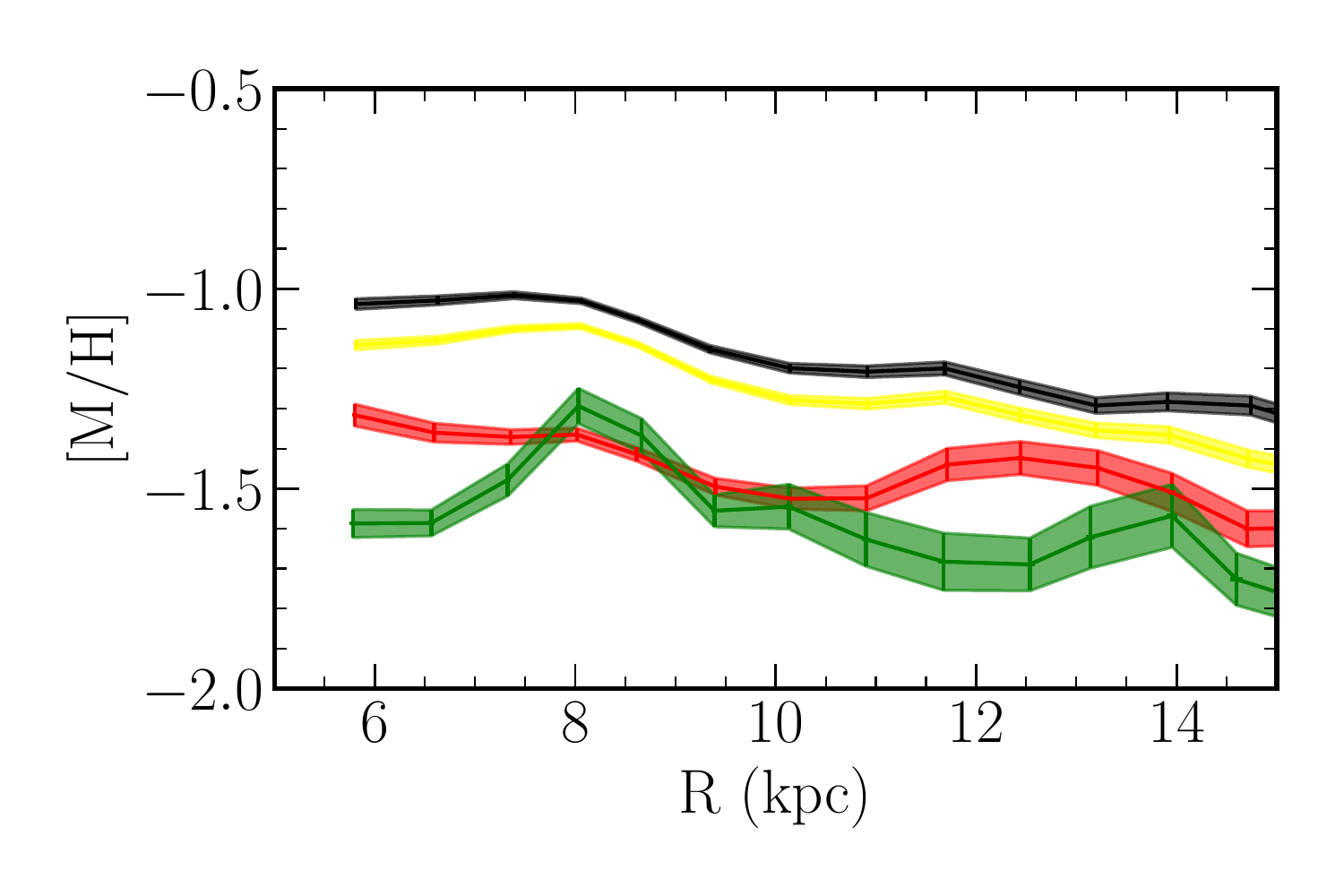}
\caption{Radial gradients for the `Gaia-Enceladus-Sausage' sample (black), the `Thamnos' sample (red), and the `other' retrograde stars (green). The criteria for selecting those stars are described in Sect.~\ref{sec:Enceladus}. The gradient for all of the retrograde stars considered simultaneously is shown in yellow.}
\label{fig:radial_gradients_GES}
\end{center}
\end{figure}

\begin{figure}
\begin{center}
\includegraphics[width=\linewidth, angle=0]{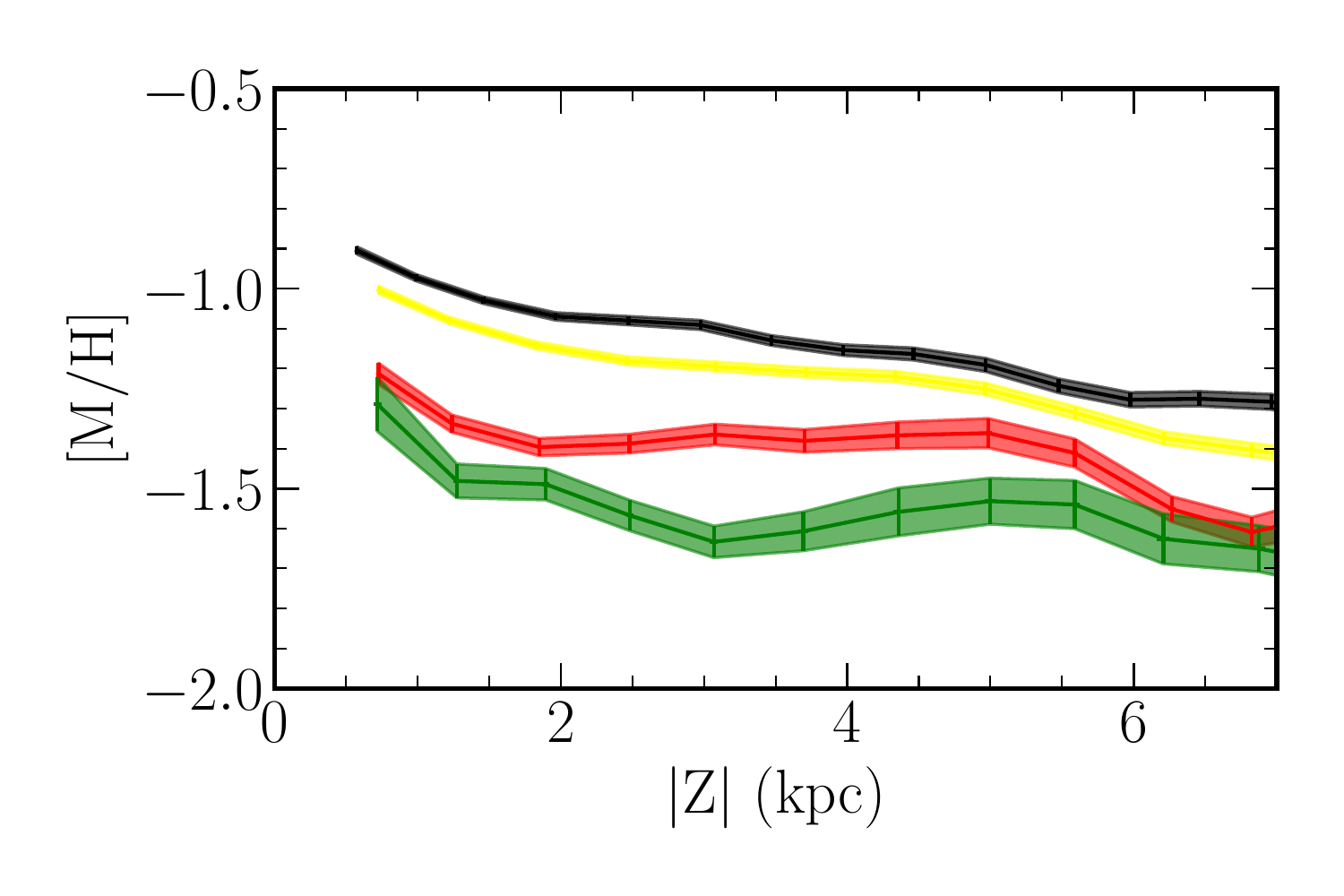}
\caption{Vertical gradients for the `Gaia-Enceladus-Sausage' sample (black), the `Thamnos' sample (red), and the 'other' retrograde sample (green). The gradient for all of the retrograde stars considered simultaneously is shown in yellow. }
\label{fig:vertical_gradients_GES}
\end{center}
\end{figure}

\begin{figure}
\begin{center}
\includegraphics[width=\linewidth, angle=0]{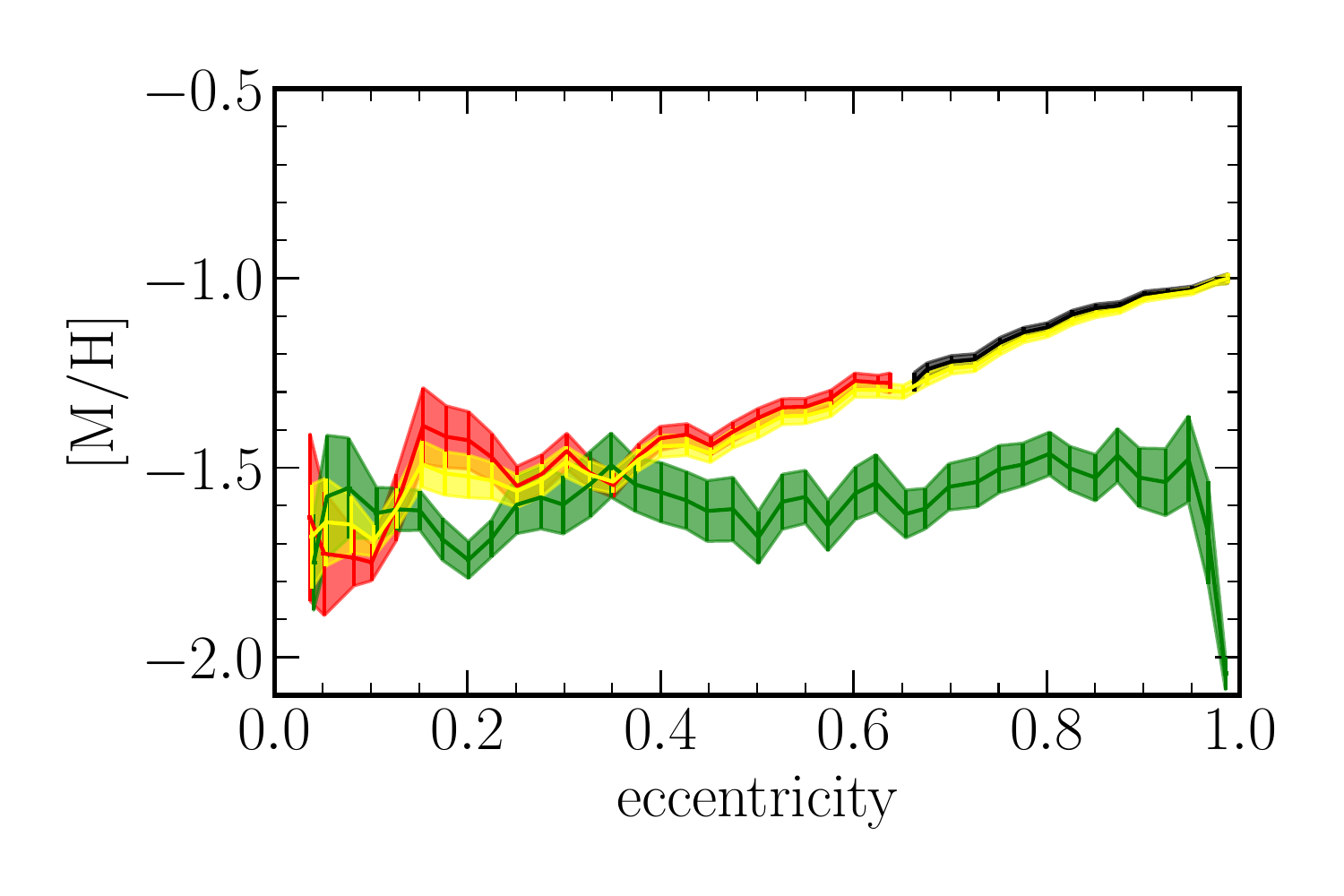}
\caption{Eccentricity gradients for the `Gaia-Enceladus-Sausage' sample (black), the `Thamnos' sample (red), and the 'other' retrograde stars (green). The gradient for all of the retrograde stars considered simultaneously is shown in yellow. }
\label{fig:enccentricity_gradients_GES}
\end{center}
\end{figure}

Figures~\ref{fig:radial_gradients_GES} and \ref{fig:vertical_gradients_GES} show the radial and vertical metallicity gradients for those three samples. 
We did not separate them into different vertical or radial ranges, as we did in Sect.~\ref{subsec:retrograde_gradients}, since we found that the gradients for the retrograde stars  are not particularly dependent on spatial cuts. 
We find that the `Gaia-Enceladus-Sausage' sample is globally more metal-rich than the `Thamnos' sample \citep[in agreement with][]{Koppelman19,Mackereth19} and that the two former samples are, in turn, more metal-rich than the `other' counter-rotating sample in our sample.

\begin{table*}
\caption{Measured radial and vertical metallicity gradients for the `Gaia-Enceladus-Sausage', `Thamnos', and `other' counter-rotating samples.}
\label{tab:Radial_Gradients_Enceladus}
\begin{center}
\begin{tabular}{cccc|ccc}
\hline \hline
Survey & Radial `Enceladus' & Radial `Thamnos' & Radial `other' & Vertical `Enceladus' & Vertical `Thamnos' & Vertical `other' \\
         & $(\dex\kpc^{-1})$ & $(\dex\kpc^{-1})$ & $(\dex\kpc^{-1})$ & $(\dex\kpc^{-1})$ & $(\dex\kpc^{-1})$ & $(\dex\kpc^{-1})$ \\ \hline
All             &  $-0.033 \pm  0.002$ & $-0.024 \pm 0.005$ & $-0.030 \pm 0.010$ &  $-0.057 \pm 0.003$ & $ -0.039 \pm 0.009$ & $ -0.030 \pm 0.010$ \\
LAMOST  &  $-0.040 \pm  0.007$ & $-0.019 \pm 0.005$ & $-0.007 \pm 0.007$ & $-0.057 \pm 0.003$ & $ -0.023 \pm 0.004$ & $ -0.021 \pm 0.006$\\
APOGEE  & $-0.015  \pm  0.003$ & $-0.021 \pm 0.011$ & $-0.012 \pm 0.011$ & $-0.077 \pm 0.007$ & $-0.108 \pm 0.017$ & $-0.029 \pm 0.010$ \\ \hline
 \end{tabular}
 \end{center}
 \end{table*}

When considering all of the surveys simultaneously, the radial and vertical metallicity gradients that we measure for `Gaia-Enceladus-Sausage', `Thamnos', and the `other' sub-samples are similar, within $1\sigma$ and $2\sigma$, respectively (see Table\,\ref{tab:Radial_Gradients_Enceladus}), yet this result should be taken with a grain of salt as it is not necessarily corroborated when considering LAMOST or APOGEE targets separately (maybe due to small number statistics or different volume coverage). 

This lack of evidence of difference in the metallicity gradients,  other than the zero-point offset,  between each of the counter-rotating sub-samples,  made us investigate  how the metallicity changed as a function of the eccentricity. Results are shown in Fig.~\ref{fig:enccentricity_gradients_GES}. Strikingly, we find that `Gaia-Enceladus-Sausage' and `Thamnos' samples follow the same trend, as opposed to the `other' subsample, which appears to show no variation of metallicity as a function of the eccentricity (a result that is also found when separating the targets per survey).  This trend seems to favour the fact that the `Thamnos' and `Gaia-Enceladus-Sausage' samples are strongly linked, yet the question remains about whether they originate from the same parent population.

The $\mgfe-\meta$ plot coming from  the APOGEE-DR16 data (Fig.~\ref{fig:CR_alphas})  shows not only that our  `Gaia-Enceladus-Sausage' and `Thamnos' samples are mixed populations, containing both high-$\alpha$ and low-$\alpha$ stars, but also that they exhibit different distributions in both the metallicity space and the $\mgfe$ space (see coloured dots and histograms).  The two sample Kolmogorov-Smirnov tests of the metallicity and $\mgfe$ between the `Gaia-Enceladus-Sausage' and `Thamnos' samples strongly suggest that the two samples cannot have been drawn from the same underlying distribution, with $p-$values always lower than $10^{-7}$, even when decomposed into radial and vertical spatial bins.

\begin{figure}
\begin{center}
\includegraphics[width=\linewidth, angle=0]{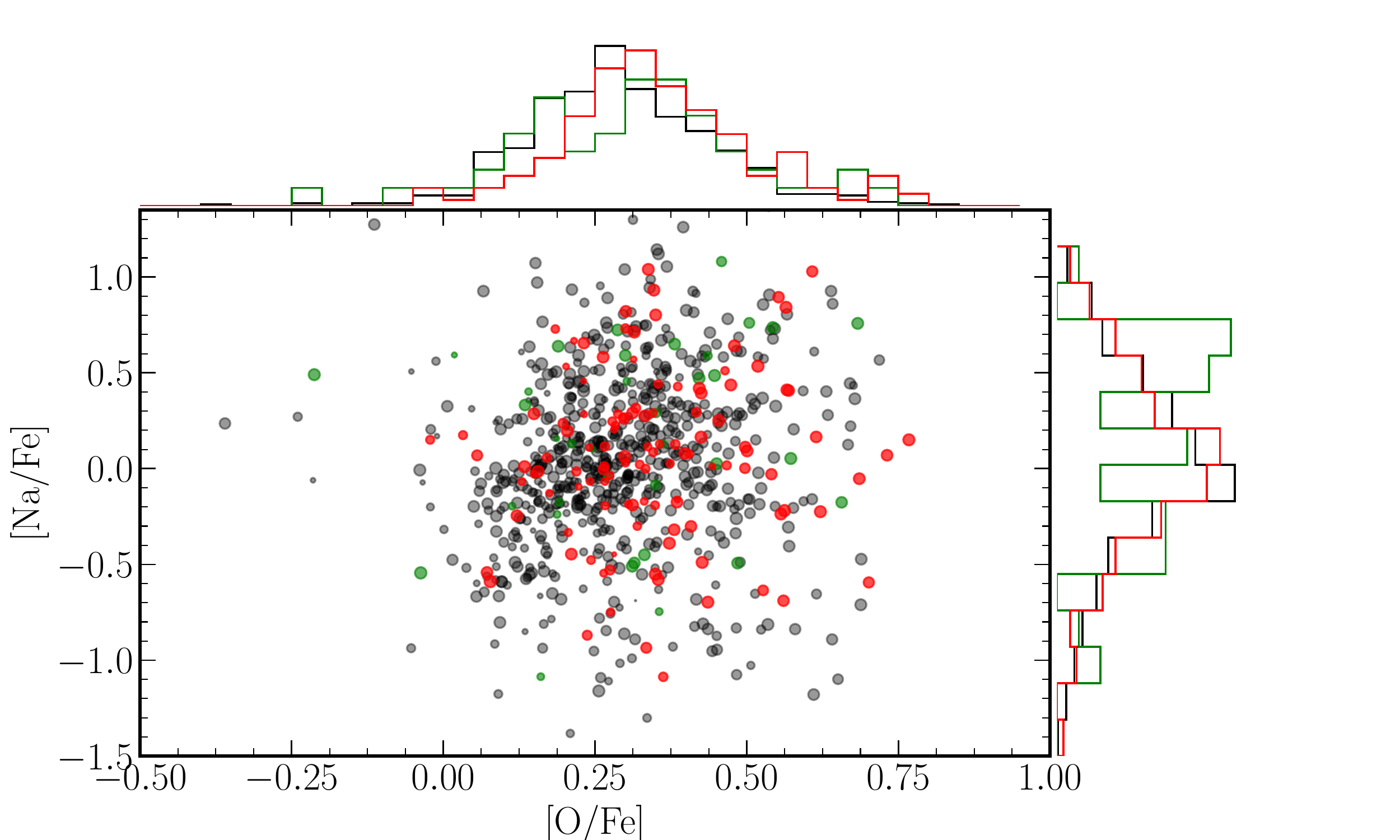}
\caption{Same as Fig.~\ref{fig:CR_alphas} but for sodium abundance as a function of oxygen for the retrograde stars in the APOGEE-DR16 catalogue. }
\label{fig:CR_Na_O}
\end{center}
\end{figure}

Finally, we also explored the possibility of having a significant amount of stars belonging to the retrograde (and peculiar on many aspects) globular cluster  $\omega$Centauri \citep[e.g.][]{Dinescu02b,Morrison09, Nissen10, Majewski12, Myeong18c}. For this purpose, we visually investigated the [Na/Fe]-[O/Fe] space for indications of an anti-correlation between the Na abundance and the O one, indicative of the one found in $\omega$Cen \citep[e.g.,][]{Johnson10, Gratton11, Marino11}. Figure~\ref{fig:CR_Na_O} shows no clear indication of anything of the sort for either one of the three sub-samples, suggesting that, globally, our sample of retrograde stars does not contain any obvious sub-population of stars stripped from the $\omega$Cen progenitor.  We note, however, as remarked by \citet{Zasowski19}, that sodium abundances in APOGEE exhibit a large scatter compared to the published uncertainties, partly due to the presence of strong telluric absorptions close to the Na lines. Therefore, firmer conclusions on this topic might require a more in-depth investigation of   the abundance patterns of APOGEE, which is beyond the scope of this paper.   

\section{Discussion and conclusions}
\label{sec:conclusions}

The merger-tree of the Milky Way is a matter of vivid debate. Despite the recent discoveries of many over-densities  in the action-energy space in the extended Solar neighbourhood, it is still rather unclear how these populations relate to each other, or to what extent they are well-mixed within the Galaxy. In this paper, we use a compiled catalogue of the metallicities and velocities of more than 4 million stars from \citet{Sanders18} to investigate the chemical gradients and the chemodynamics of the counter-rotating stars, while also performing similar analyses for the prograde stars. This parallel analysis allowed us to highlight the differences between the two populations and also to confirm that our analysis method returned sound results.

\bigskip 

Our main results can be summarised as follows: 
\begin{itemize}
\item
We determined that stars on prograde orbits show a $\vphi-\meta$ relation that flattens -- and perhaps even reverses as a function of distance from the plane. To our knowledge, this trend has not yet been reported in the literature.

\item
We found a super-solar metallicity counter-rotating population over a large range of radii and distances from the plane ($|Z|\sim5-6\kpc$ and $R\sim15\kpc$).  

\item
Retrograde stars exhibit vertical and radial metallicity gradients that are similar at all probed radii and distances from the Galactic plane, respectively ($\partial \meta / \partial R \approx -0.04\dex\kpc^{-1}$ and $\partial \meta / \partial |Z| \approx -0.06\dex\kpc^{-1}$).

\item 
The correlation between $\vphi$ and the metallicity for the retrograde stars also seem to be similar everywhere in the Galaxy. 

\item 
Samples roughly probing `Gaia-Enceladus-Sausage' and `Thamnos' (or Sequoia's low-energy tail)  show similar metallicity  gradients but different metallicity and $\afe$ distributions. 
Despite being truncated and likely contaminated samples, 
we find that they are following the exact same sequence of metallicity versus eccentricity. Other counter-rotating stars with lower $\vphi$ do not follow the same sequence.

\item
Counter-rotating stars do not show any striking anti-correlations in [Na/Fe] versus [O/Fe] chemical space, suggesting that our sample does not contain many $\omega$Cen stars. 
\end{itemize}

We investigated potential biases due to parallax zero-point offsets or the inhomogeneity of the survey metallicities in order to make sure that the results of our analysis are robust. Globally, a correction of the parallax zero point  affects mainly the retrograde stars, increasing their velocities and, hence, decreasing the sample-size of this population. Our conclusions, however, remain robust as they do not rely on the absolute densities of the populations. The different zero points in metallicity between the surveys do not alter our conclusions either. Indeed, where, for example, the absolute values of the metallicity gradients might differ if a specific survey is used instead of another, the global trends remain unchanged. That said, we stress that additional biases might still exist due to the different quality cuts that we applied to our input catalogue. These are difficult to quantify (e.g. the effect of filtering for the stars with large uncertainty in metallicity), still, we present a brief discussion in Appendix~\ref{sec:bias_cuts}. 

\bigskip

From the points above, the following picture can be drawn. 
The flattening (or inversion of sign, from positive to negative) of the correlation of $\vphi$ with metallicity for the prograde stars far from the plane may suggest, in agreement with other studies, that the inner halo and the thick disc far from the plane may be populations that share a common past. 
Indeed, it is believed that mono-age disc populations have intrinsically negative $\vphi-\meta$ correlations \citep[][]{Minchev19}. Hence, going from a positive correlation for the thick disc stars close to the plane to a negative  (or flat) one for the stars at large distances from it, where the probed population is a mixture of thick disc and halo, would indicate that the targeted population has a smaller age range.

One point that is worth mentioning here relates to recent results, for example, from  \citet{Haywood18, Fernandez-Alvar19b, Myeong19, DiMatteo19, Gallart19, Belokurov20},  who revived the idea that the low-angular momentum thick disc might be made up of stars from the proto-disc that are now part of the inner halo after having been heated by a major past accretion \citep[see also,][]{Gilmore02,Wyse06,Kordopatis13a}. In particular, using the same input catalogue as we have in this work, \citet{Belokurov20} identified  a small, yet non-negligible, prograde population of intermediate metallicity ($\meta\gtrsim-0.7$)  with low angular momentum ($\vphi\lesssim100\kms$), which they have dubbed `the Splash'. These stars are found to be slightly younger than the accreted ones from the last major merger (i.e. Gaia-Enceladus-Sausage)  and with a different star-formation history than the latter. These aspects led these authors to suggest that the Splash stars were born locally in the protodisc and have had their orbits altered by this massive accretion \citep[see, however,][for an alternative explanation]{Amarante20}. 
Our analysis, in selecting each time only the prograde or the retrograde stars, unavoidably dilutes the signature of the Splash with either the thick disc or the halo and Gaia-Enceladus-Sausage targets. However, investigating Figs.~\ref{fig:vphi_metal_sun} to \ref{fig:vphi_metal_outer}, at around $100\lesssim \vphi \lesssim 150\kms$ \citep[where ][suggest that there is the overlap between the Splash and the disc]{Belokurov20}, we do not see any peculiarities in the trend other than the well-known change of behaviour in the $\vphi-\meta$ space, going from an anti-correlation to a positive correlation.

\bigskip

As far as the counter-rotating stars are concerned, we find that  more retrograde targets are also, on average,  more metal-poor. The lack of spatial changes in the metallicity gradients, or in the $\vphi-\meta$ correlations, within the probed volume ($5.2<R<11.2\kpc$ and $|Z|<10\kpc$),  indicate that the counter-rotating stars are remarkably well-mixed.  This is in line with, for example, \citet{Deason18, Helmi20} \citep[and with hints of this already discussed in ][]{Watkins09, Deason13} suggesting that the orbital turning points (i.e. shells) of Gaia-Enceladus-Sausage are at a distance of $\sim20\kpc$, that is, outside the volume that we investigate here. Our results expand upon this, as they show that none of the shells of Gaia-Enceladus-Sausage, Thamnos, Sequoia, or any other accreted population that could compose the retrograde stars, is within $10-12\kpc$ from the Sun.

Amongst the counter-rotating stars, 
the samples we dub `Thamnos' and `Gaia-Enceladus-Sausage' seem to compose two different populations that are, nevertheless, also somehow linked. 
Indeed, we find that the `Thamnos' sample is more metal-poor than the `Gaia-Enceladus-Sausage' one and on more circular orbits. However,  it is striking to observe that they follow the same trend in the metallicity-eccentricity space. This trend cannot be explained simply by the presence of intrinsic metallicity gradients within a common progenitor of Gaia-Enceladus-Sausage and Thamnos/Sequoia. Indeed, despite the initial mass estimates of Gaia-Enceladus-Sausage found in the literature,  ranging from $6\cdot10^8$ to $5\cdot10^9$\,M$_\odot$ \citep[e.g.][]{Helmi18, Myeong19, Kruijssen19, Mackereth19,Fattahi19, Feuillet20}, which could lead to such internal radial metallicity gradients, it is unclear how a counter-rotating accretion with an angle of $\sim 30\deg$ \citep[see][]{Helmi20}  would deposit its most metal-rich stars, which would otherwise be expected to be present at the innermost regions of the progenitor at the highest eccentricities.

\bigskip

We believe that the key to understanding the effect of the past accretions on the properties of the thick disc  is the super-solar metallicity  counter-rotating population that we identified in our sample. These stars, if proven to really exist, could have been  formed from local gas at the moment of the accretion with the Gaia-Enceladus-Sausage progenitor \citep[see ][ for star formation activity within galactic outflows]{Maiolino17, Gallagher19}.  This population would therefore offer us an undeniable sample of locally born retrograde stars in order to precisely date and weigh  this merger \citep[see,][]{Grand20}. 
The Gaia data release 3, which will become public at the end of 2020,  will first confirm whether these peculiar stars are indeed counter-rotating. Then, future large spectroscopic surveys such as WEAVE \citep{Dalton18} or 4MOST \citep{deJong19} will  allow us to verify the chemical composition of these stars, investigate them in more dimensions of the chemical space, and, eventually, to better understand their link with the rest of the galactic populations.

\section*{Acknowledgments}
We thank the anonymous referee for their useful feedback that helped improving the quality of this paper. 
This work has made use of data from the European Space Agency (ESA)
mission {\it Gaia} (\url{https://www.cosmos.esa.int/gaia}), processed by
the {\it Gaia} Data Processing and Analysis Consortium (DPAC,
\url{https://www.cosmos.esa.int/web/gaia/dpac/consortium}). Funding
for the DPAC has been provided by national institutions, in particular
the institutions participating in the {\it Gaia} Multilateral Agreement.
This research made use of Astropy,\footnote{http://www.astropy.org} a community-developed core Python package for Astronomy \citep{astropy:2013, astropy:2018}.
This work was supported by the Programme National Cosmology et Galaxies (PNCG) of CNRS/INSU with INP and IN2P3, co-funded by CEA and CNES and ANR 14-CE33-014-01. GK acknowledges support from OCA and Lagrange BQR.
Jason Sanders and Patrick de Laverny are warmly thanked for useful discussions regarding the catalogue used in this work and the quality criteria to apply. 

\bibliographystyle{aa}
\def\aj{AJ}\def\apj{ApJ}\def\apjl{ApJL}\def\araa{ARA\&A}\def\apss{Ap\&SS}
\def\mnras{MNRAS}\def\aap{A\&A}\def\nat{Nature}
\def\nar{New Astron. Rev.}

\bibliography{../master_bib}

\begin{appendix}

\section{Change in the results when applying a parallax offset}
\label{sec:parallax_offset}
\begin{figure}
\begin{center}
\includegraphics[width=\linewidth, angle=0]{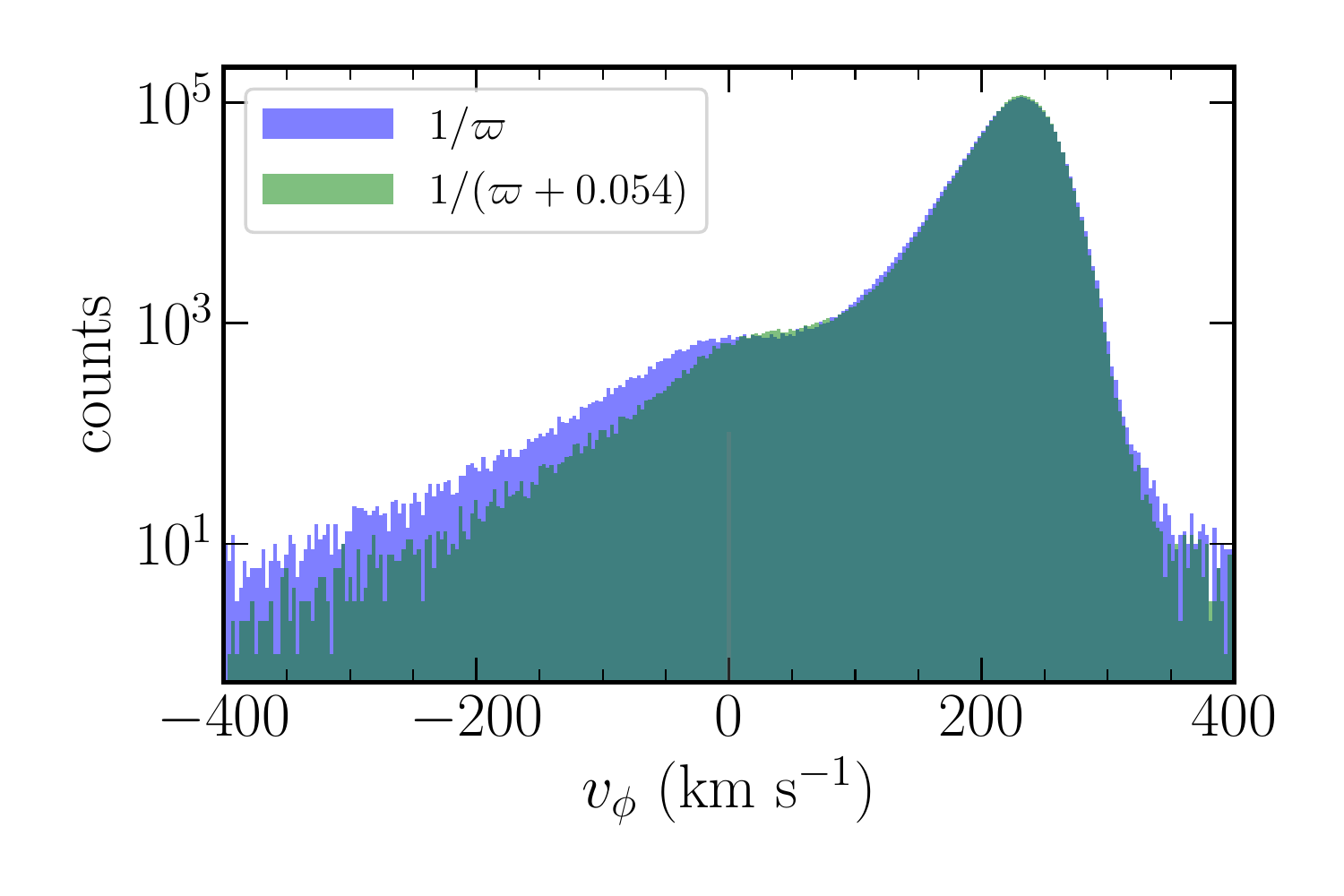}
\caption{ Azimuthal velocity distributions, in logarithmic scale,  for velocities computed with (green histogram) and without (blue histogram) zero-point parallax correction.  }
\label{fig:Vphi_DF}
\end{center}
\end{figure}

Several studies, including the ones from Gaia's Data Processing Analysis Consortium (DPAC)\footnote{\url{https://gea.esac.esa.int/archive/documentation/GDR2/}} have reported a zero-point shift in the parallax values of GDR2, in the sense that Gaia parallaxes would need to be increased.  
Using quasars, \citet{Lindegren18} reported the mean zero-point  to be $\delta \varpi=-0.03$\mas, with variations depending on the sky position, the target's magnitude and colour \citep{Arenou18}. Similarly, \citet{Graczyk19} and \citet{Schonrich19}, using binaries and all of the stellar sample with radial velocities, respectively, reported it to be $\delta \varpi=-0.054$\mas.  A similar zero-point offset was determined towards the Kepler and K2 fields by \citet{Zinn19} and \citet{Khan19} using RGB and RC asteroseismic targets.
An overestimation in distances leads to an under-estimation of the azimuthal velocities, hence, in a decrease of the number of retrograde stars as shown in Fig.~\ref{fig:Vphi_DF} \citep[see also discussion in][ and references therein for the effect of over-estimated distances on the velocities of the stars]{Schonrich11}.

\begin{figure}
\begin{center}
\includegraphics[width=\linewidth, angle=0]{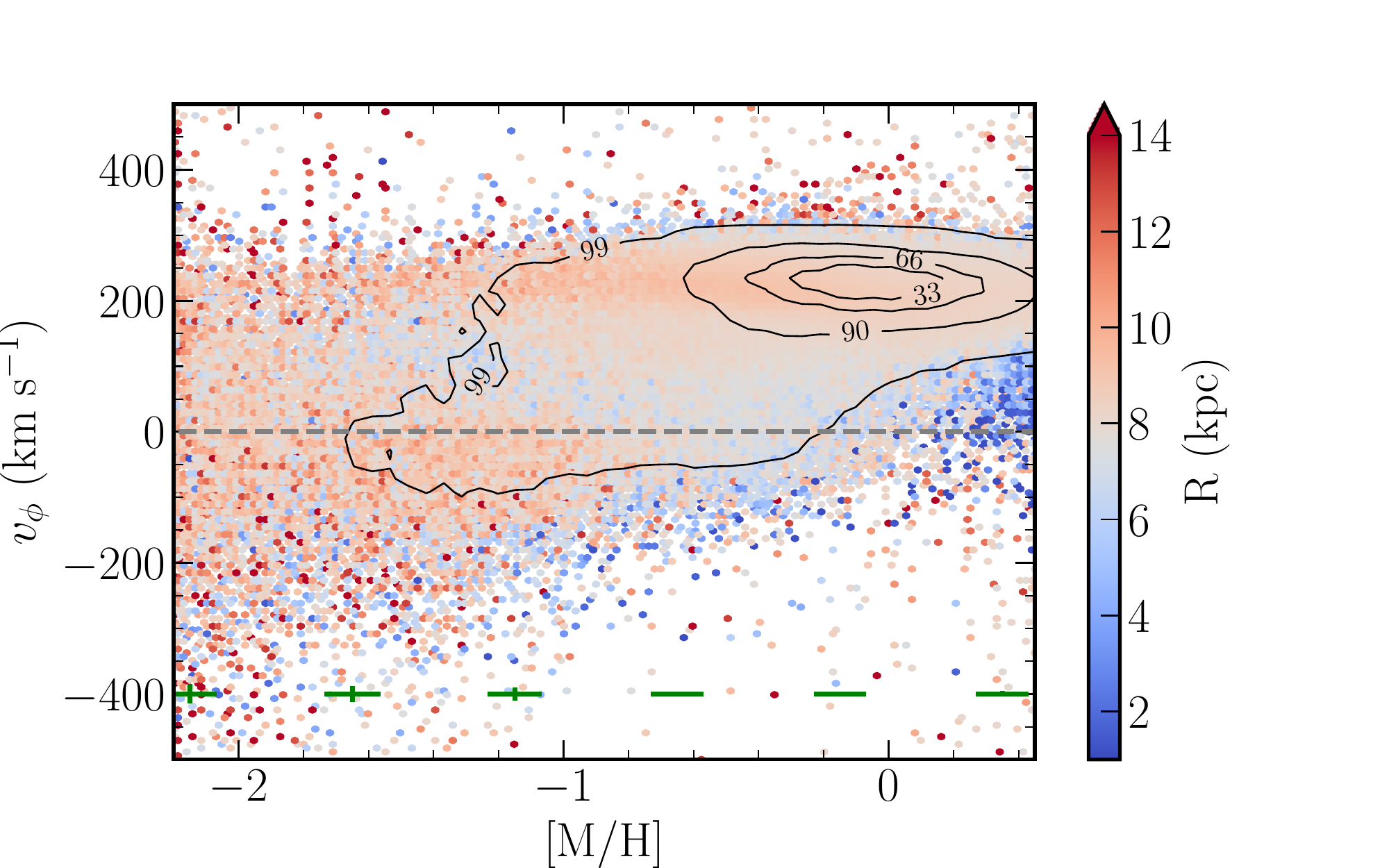}\\
\includegraphics[width=\linewidth, angle=0]{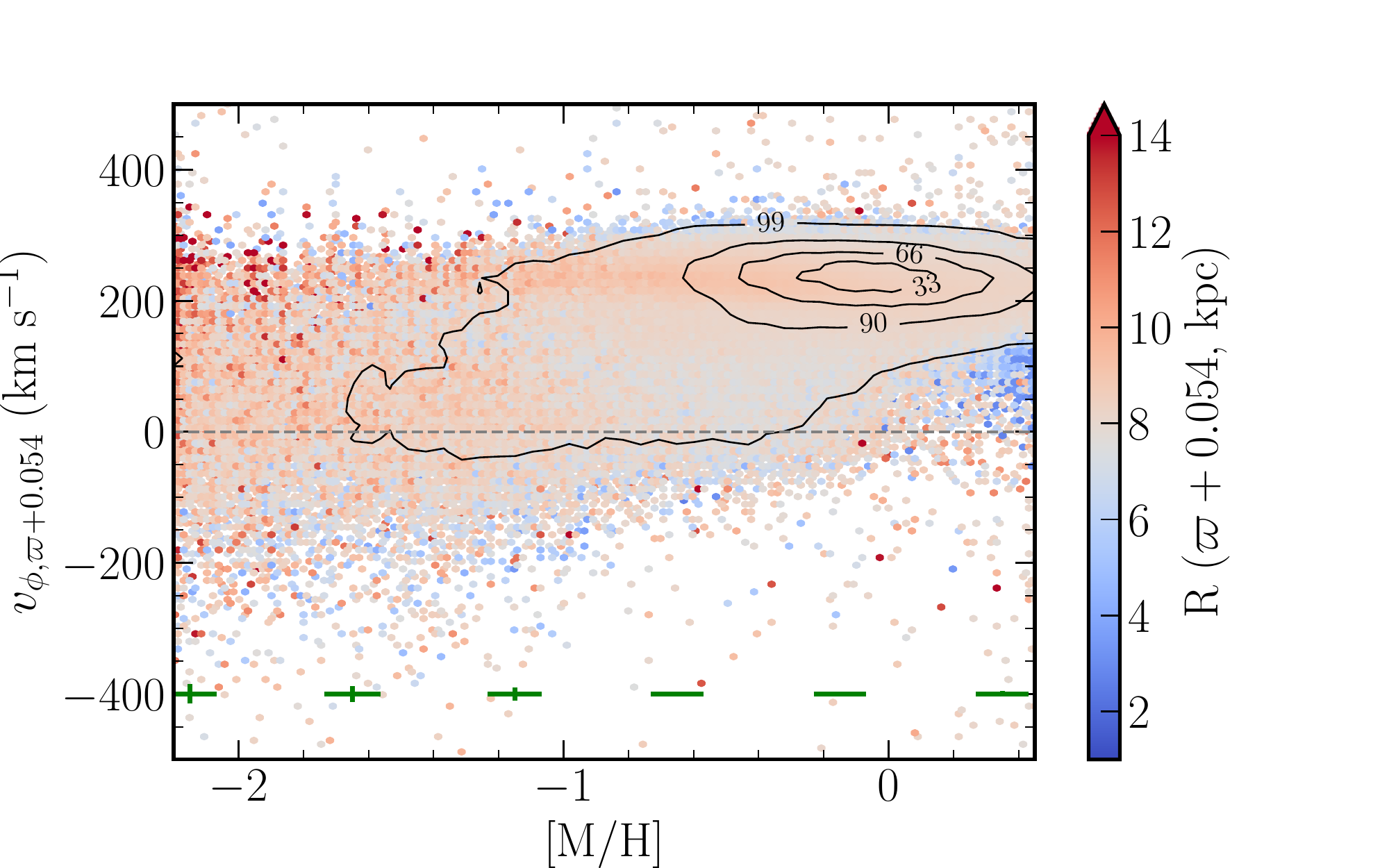}
\caption{$\vphi$ vs $\meta$ for the all of the stars in our sample, without (top) and with (bottom) a zero-point correction of $0.054$\,mas on the parallaxes. The uncertainties in $\vphi$ and $\meta$ for stars in bins of $\meta$ are shown in green at the bottom of the plots. The colour code corresponds to the mean Galactocentric radius of the stars with and without the zero-point correction. Contour lines enclose 33, 66\%, 90\%,\ and 99\%\ of the sample. Dashed grey line denotes the region where $\vphi=0\kms$. }
\label{fig:metal_vphi_all_R}
\end{center}
\end{figure}

\begin{figure}
\begin{center}
\includegraphics[width=\linewidth, angle=0]{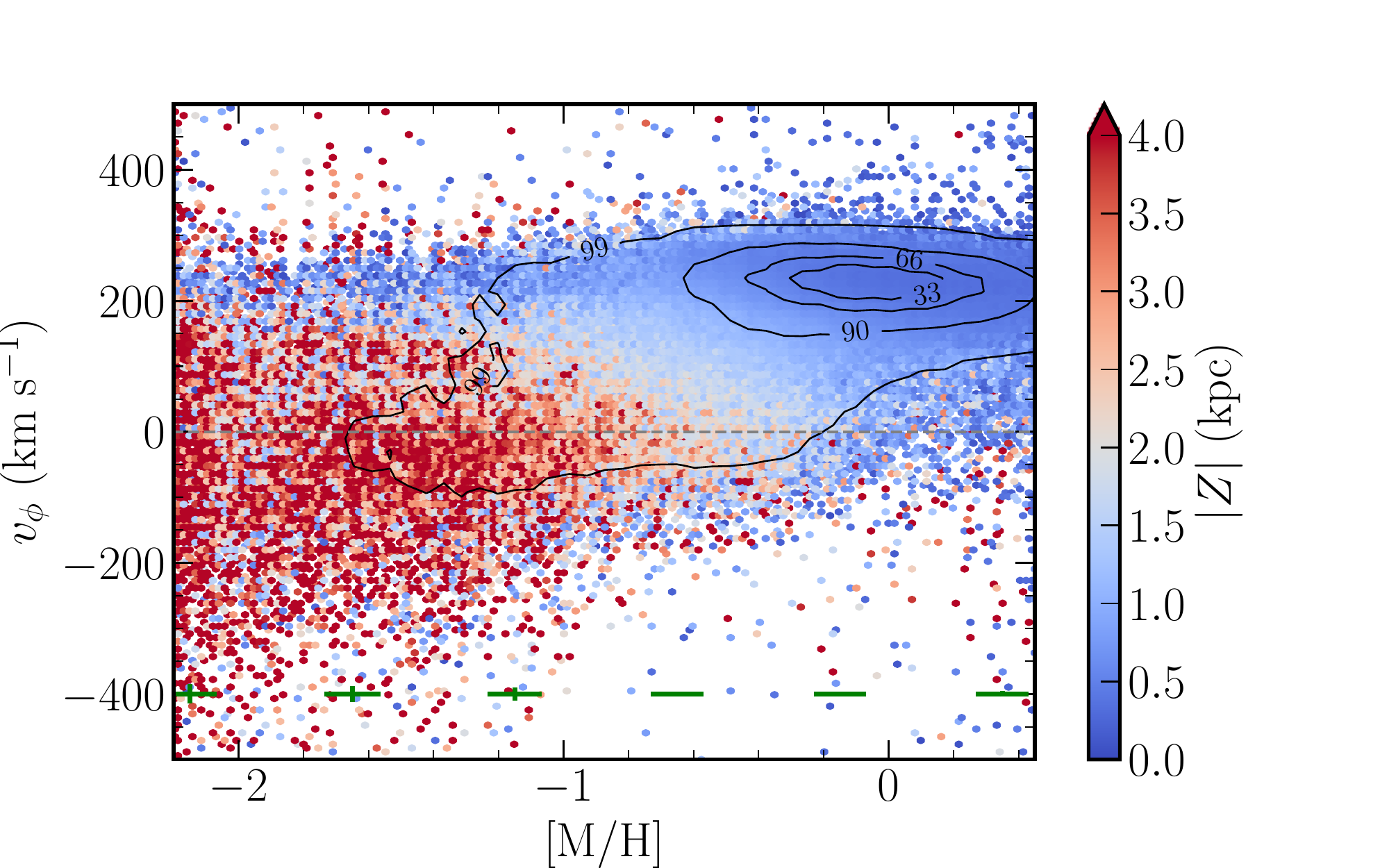}\\
\includegraphics[width=\linewidth, angle=0]{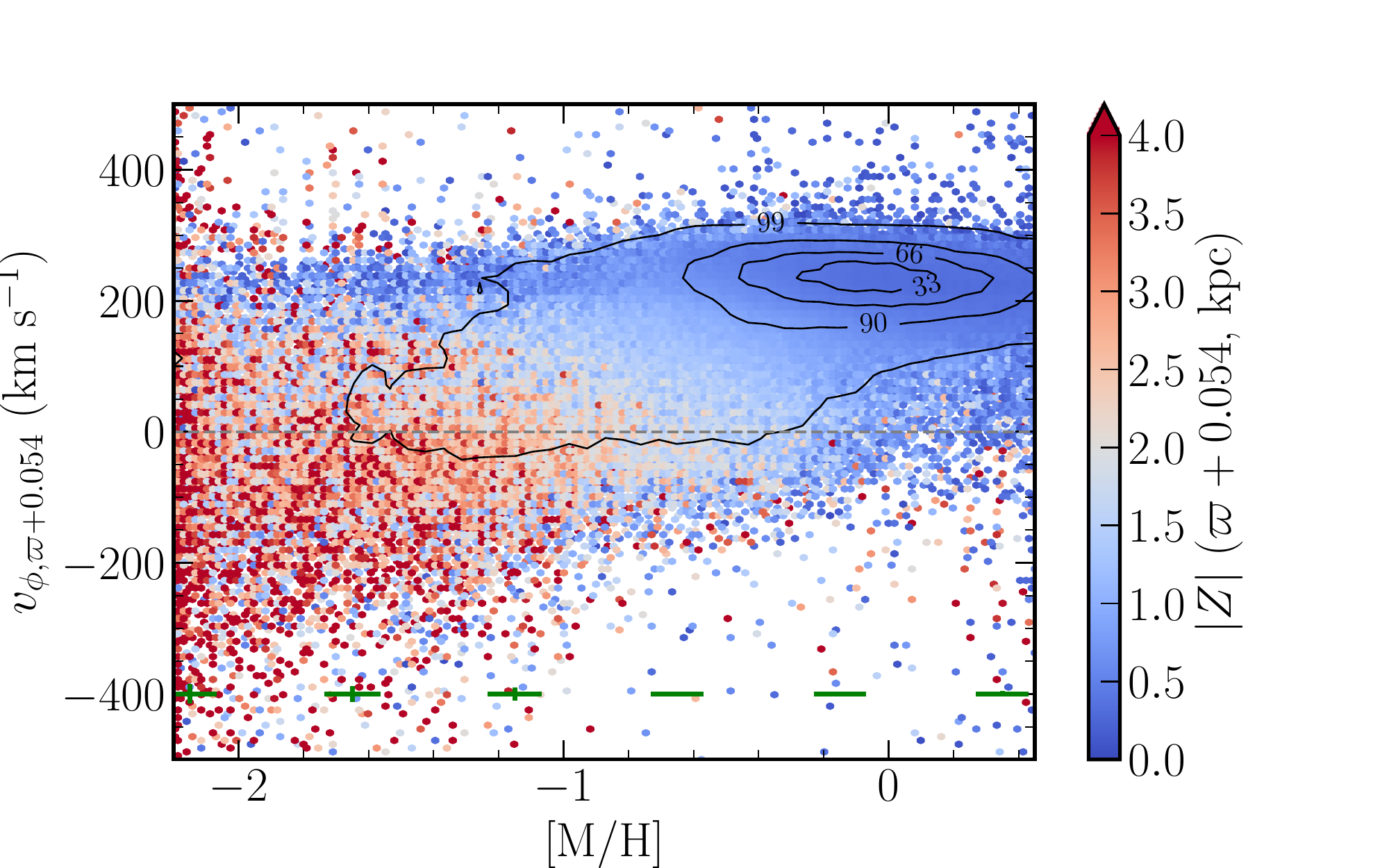}
\caption{Same as Fig.~\ref{fig:metal_vphi_all_R}, with a colour code corresponding to the mean absolute distance from the Galactic plane, $|Z|$.   }
\label{fig:metal_vphi_all_Z}
\end{center}
\end{figure}

Figures~\ref{fig:metal_vphi_all_R}, \ref{fig:metal_vphi_all_Z} show the way the azimuthal velocity and  the $R-Z$ positions change when taking into account a correction of $0.054$\,mas. The velocities of prograde stars are merely affected, with shifts smaller than $20\kms$, whereas $\sim50\%$  of the retrograde stars have shifts that are less than $40\kms$ and $\sim85\%$  less than $100\kms$.

\begin{table}
\caption{Measured Galactocentric radial gradients in the $R=[5-15]\kpc$ range with zero-point correction.}
\label{tab:Radial_Gradients_zeropoint}
\begin{center}
\begin{tabular}{ccc}
\hline \hline
$|Z|$-range & Prograde & Retrograde \\
(kpc) & (dex/kpc) & (dex/kpc) \\ \hline
$[0.2-1.0]$&  $-0.042\pm 0.004$ & $-0.065\pm 0.012$ \\ 
$[1.0-2.0]$& $-0.003 \pm  0.004$ & $-0.047 \pm 0.008$ \\ 
$[2.0-4.5]$& $-0.010 \pm 0.005$ & $-0.03\pm 0.007$\\ \hline
 \end{tabular}
 
 \end{center}
 \end{table}

 \begin{table}
\caption{Measured vertical gradients relative to the Galactic plane in the range $|Z|=[0-5]\kpc$}
\label{tab:Vertical_Gradients_zeropoint}
\begin{center}
\begin{tabular}{ccc}
\hline \hline
$R$-range & Prograde & Retrograde \\
(kpc) & (dex/kpc) & (dex/kpc) \\ \hline
$[5.2-7.2]$ & $-0.302  \pm 0.007$ & $-0.108 \pm  0.007$ \\
$[7.2-9.2]$& $-0.228 \pm  0.009$ & $-0.122 \pm 0.009$ \\
$[9.2-11.2]$ & $-0.195  \pm 0.009$ & $-0.065  \pm 0.008$ \\ \hline

 \end{tabular}
 \end{center}
 \end{table}

Finally, Tables~\ref{tab:Radial_Gradients_zeropoint} and \ref{tab:Vertical_Gradients_zeropoint} show the radial and vertical metallicity gradients, respectively, when correcting for a zero-point parallax offset of $0.054$\,mas.


\section{Comparison between repeats in several surveys}
\label{sec:intersurveyComparison}
\begin{figure*}
\begin{center}
\includegraphics[width=0.82\linewidth, angle=0]{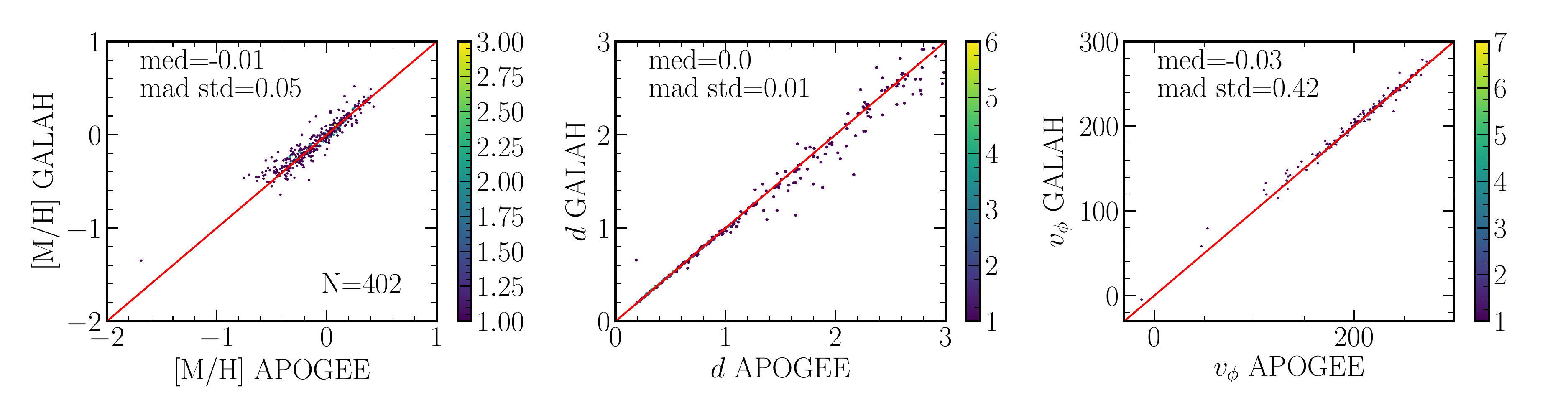}\\
\includegraphics[width=0.82\linewidth, angle=0]{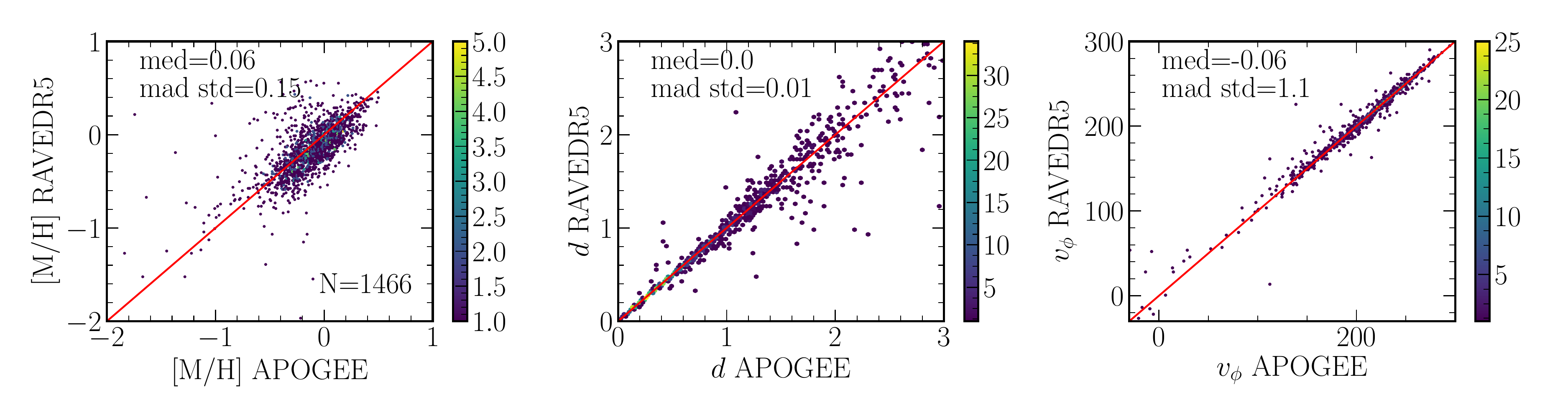}\\
\includegraphics[width=0.82\linewidth, angle=0]{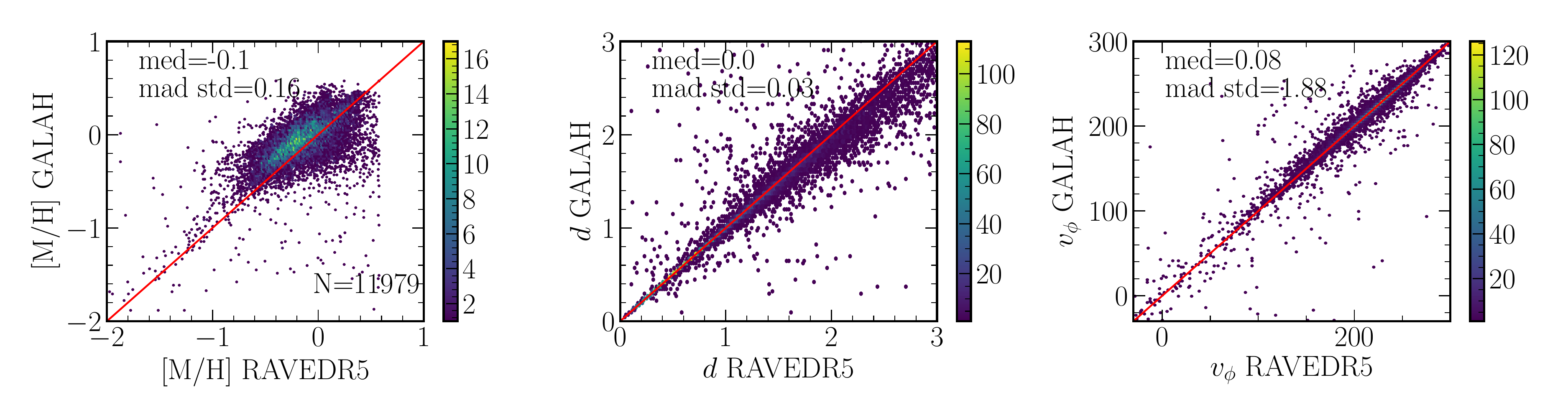}\\
\includegraphics[width=0.82\linewidth, angle=0]{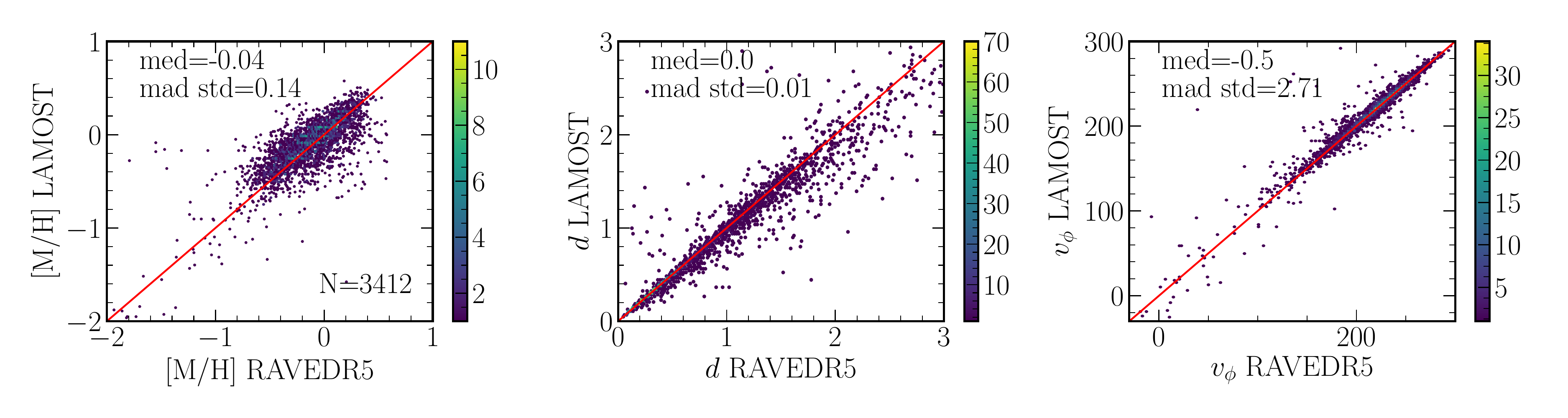}\\
\includegraphics[width=0.82\linewidth, angle=0]{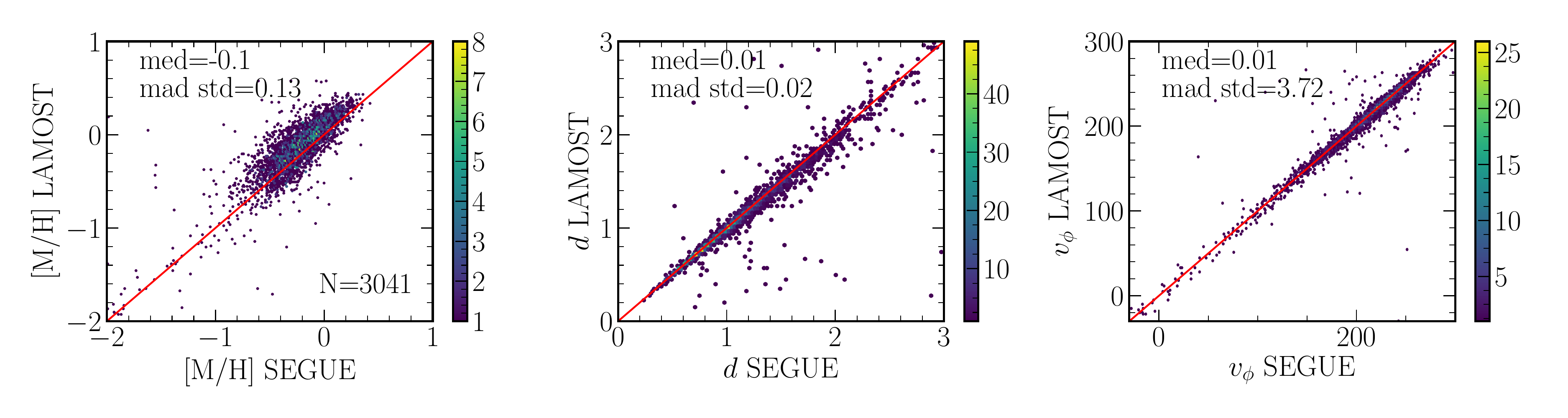}
\caption{2D histograms showing the offsets for duplicate targets between APOGEE, RAVE, GALAH, LAMOST, and SEGUE when at least 100 targets were in common. The colour-code corresponds to the number of stars contained inside each hexbin. Median offsets ($x-$axis - $y-$axis) and robust standard deviations (calculated correcting the Median Absolute Deviation by a factor of 1.4826) are reported at the top-left corners of each panel. First column shows comparison for line-of-sight distances, second column for metallicities and third column for azimuthal velocities.}
\label{fig:survey_offsets}
\end{center}
\end{figure*}

We show in Fig.~\ref{fig:survey_offsets}, the comparison between the metallicity, the distance, and the azimuthal velocity, as derived by \citet{Sanders18}, for the repeated inter-survey stars. A very good agreement is obtained for distances and $\vphi$ (largest median offsets are $10\pm20$\,pc~and  $0.5\pm2.7\kms$, respectively), whereas metallicity offsets are smaller or equal to $0.10\pm0.16 \dex$ (largest median offset found is between RAVE and GALAH). Since we keep only the stars that exhibit a very good astrometry and a small uncertainty in metallicity (see Sect.~\ref{sec:dataset}), this offset in \meta~ mostly translates the offset in metallicity derived spectroscopically by each survey.  Eventually, we kept only one entry for those repeated stars, with the following order of preference: APOGEE, GALAH, GES, RAVE, LAMOST, and SEGUE.

\section{Effects on our analysis of the different quality cuts applied}
\label{sec:bias_cuts}

The quality cuts we impose on our input dataset might introduce some undesired biases in our analysis. These are irrelevant when discussing, for example, the existence of counter-rotating super-solar metallicity stars, yet they might affect the definition of the retro-grade and prograde samples, or the measurements of the metallicity gradients and the $\vphi-\meta$ correlations. Below, we offer a qualitative comment on how these cuts affect our analysis.
\begin{enumerate}
\item
Cuts in astrometric quality to remove potential binaries (RUWE and astrometric excess noise): We do not expect that this cut introduces any biases, nor is it affected by zero-point offsets. 
\item
 Cuts to remove stars with poor parallax measurements ($\sigma\varpi/\varpi>0.1$): This cut tends to remove preferentially distant stars. At a given distance, no metallicity bias should be introduced, hence, it should not be when investigating specific regions of the Galaxy either.  A parallax zero-point offset would tend to change (increase, in this case) the size of our working sample.   
\item
 Cuts to remove cool and hot stars: This cut potentially removes metal-rich dwarfs (hot stars) and metal-poor giants (cool stars). The relative proportion of metal-rich over metal-poor depends on the selection function of the different surveys. 
\item
 Cuts to remove stars with large metallicity uncertainty: This cut would tend to remove preferentially metal-poor stars (due to the lack of spectral signatures to derive the stellar parameters). Yet, each survey observes in different wavelength ranges and different resolutions. Thus, whereas this cut removes a substantial part of SEGUE (32 per cent), LAMOST (11 percent), and RAVE (2 percent), it removes less than 0.1 per cent of APOGEE, GALAH, and Gaia-ESO Survey stars. 
\item
 Cuts to remove stars with large azimuthal velocity uncertainty: What impacts the azimuthal velocity uncertainty is mainly the parallax uncertainty. The same type of biases as the ones presented in cut number 2, shown above, is therefore expected.
\end{enumerate}

\end{appendix}

\end{document}